\documentclass[fleqn,usenatbib]{mnras}

\usepackage{newtxtext,newtxmath}

\usepackage[T1]{fontenc}

\DeclareRobustCommand{\VAN}[3]{#2}
\let\VANthebibliography\thebibliography
\def\thebibliography{\DeclareRobustCommand{\VAN}[3]{##3}\VANthebibliography}

\usepackage{graphicx}	
\usepackage{amsmath}	

\usepackage{amssymb}	
\usepackage{ulem}
\usepackage[dvipsnames]{xcolor}
\usepackage{subfigure}
\usepackage{hyperref}

\title[Modeling biases from constant $M_{\star}/L$ assumption]{Modeling biases from constant stellar mass-to-light ratio assumption in galaxy dynamics and strong lensing} 

\author[Liang et al.]{
Yan Liang$^{1}$\thanks{E-mail: liangy19@mails.tsinghua.edu.cn},
Dandan Xu$^{1}$\thanks{E-mail: dandanxu@tsinghua.edu.cn},
Dominique Sluse$^{2}$,
Alessandro Sonnenfeld$^{3}$
and Yiping Shu$^{4}$ \\ 
$^{1}$ Department of Astronomy, Tsinghua University, Beijing, 100084, China\\
$^{2}$ STAR Institute, University of Li\`ege, Quartier Agora - All\'ee du six Ao\^ut, 19c B-4000 Li\`ege, Belgium\\
$^{3}$ Department of Astronomy, Shanghai Jiaotong University, Shanghai, 200240, China \\ 
$^{4}$ Purple Mountain Observatory, Chinese Academy of Sciences, Nanjing, 210023, People's Republic of China \\
}

\date{Accepted XXX. Received YYY; in original form ZZZ}

\pubyear{2021}

\begin{document}
\label{firstpage}
\pagerange{\pageref{firstpage}--\pageref{lastpage}}
\maketitle

\begin{abstract}
A constant stellar mass-to-light ratio $M_\star/L$ has been widely-used in studies of galaxy dynamics and strong lensing, which aim at disentangling the mass distributions of dark matter and baryons. However, systematic biases arising from constant $M_\star/L$ assumption have not been fully quantified. In this work, we take massive early-type galaxies from the TNG100 simulation to investigate possible systematic biases in the inferences due to a constant $M_\star/L$ assumption. We construct two-component matter density models, where one component describes the dark matter, the other for the stars, which is made to follow the light profile by assuming a constant $M_\star/L$. We fit the two-component model directly to the {\it total} matter density distributions of simulated galaxies to eliminate systematics coming from other model assumptions. We find that galaxies generally have more centrally-concentrated stellar mass profile than their light distribution. Given the light profiles adopted (i.e., single- and double-S{\'e}rsic profiles), the assumption of a constant $M_\star/L$ would artificially break the model degeneracy between baryons and dark matter {\it for non-constant} $M_\star/L$ systems. For such systems, without knowing the true $M_{\star}/L$ but assuming a constant ratio, the two-component modeling procedure tend to generally overestimate $M_{\star}/L$ by $30\%-50\%$, and underestimate the central dark matter fraction $f_{\rm DM}$ by $\sim 20\%$ on average.

\end{abstract}




\section{Introduction}

Within the dark matter cosmology framework, galaxies composed of stars and gas (baryons) live in dark matter halos. One primary goal of galaxy dynamical studies, often in combination with stellar population synthesis (SPS) and sometimes gravitational lensing, is to correctly disentangle the different contributions from the dark and the baryonic matter inside a galaxy. On the dark matter side, the mass fraction and density slope in the inner region of a galaxy (as well as their redshift evolution) may provide us with crucial information on the merger history of a galaxy (e.g., \citealt{BoylanKolchin_2005MNRAS_DissipationlessMerger-FPevolution,Hilz_2013MNRAS_MinorMerger-GalaxyShape,Tortora_2014MNRAS_ETG_DMevolution_z0.8}) as well as the dynamical influence of baryons (e.g., \citealt{Blumenthal_1986ApJ_DM_AC,Gnedin_2004ApJ_DM_ACmodel}), in particular in the presence of various feedback mechanisms (e.g., \citealt{Navarro_1997MNRAS_SNoutflow-DMcore,Duffy_2010MNRAS_Baryon_DM,Governato_2012MNRAS_CuspyNoMore,Lovell_2018MNRAS_TNG_fDM}). This is because during these processes, the dark matter halo will respond accordingly, altering its phase-space distribution. Observationally, however, the exact amount and shape of the inner dark matter halo are difficult to measure to high accuracy, largely because dark matter can only be probed indirectly, with the help of gravitational tracers and in combination of carefully deducting the baryonic contribution. This has substantially restrained our understanding on and solutions to the small-scale controversies of the standard cold dark matter theory, as inferred by observations (see \citealt{Weinberg_2015PNAS_CDMcontroversies-onSmallScale} and \citealt{Bullock_2017ARA&A_SmallScaleChallenge_LambdaCDM} for general review). On the baryonic side, although the stellar component can be directly probed by light, correctly converting luminosity $L$ to stellar mass $M_{\star}$ relies on knowing (the spatial distribution of) the mass-to-light ratio $M_{\star}/L$. The stellar mass-to-light ratio $M_{\star}/L$ depends on the age and metallicity of the underlying stellar population, as well as the initial mass function (IMF). In particular, the IMF describes the stellar mass distribution for a population at birth, and is an important fundamental property of a galaxy. A population IMF has been shown to depend on the turbulent environment of the interstellar medium (e.g., \citealt{Padoan_2002ApJ_TurbulentFragment-IMF,Hopkins_2013MNRAS_TurbulentFragment,Chabrier_2014ApJ_IMF_StarburstEnvironment}). While a galaxy-wide IMF has also been shown to correlate with properties such as the intensity of star formation, effective pressure and metallicity (e.g., \citealt{Conroy_2012ApJ_ETG_IMF_AbsorptionLine_II,Jerabkova_2018_A&A_IMF-Z&SFR,ZhouShuang_2019MNRAS_MaNGA_IMF_Variation}), essentially reflecting the stellar assembly environment and history of galaxies.   

Various modeling approaches have been developed and the exact implementation depends on the quality and richness of available observational data, which, under different circumstances, may suffer from different limitations and complications. For example, when high $S/N$ and high resolution spectroscopic data are available, IMF-sensitive absorption lines or features in a galaxy spectrum can be used to reveal the true IMF (as well as the stellar mass). Through this approach, many studies support that massive galaxies in the nearby Universe have Salpeter-like IMFs (\citealt{Salpeter_1955ApJ_IMF}), which are more bottom-heavy (i.e., a larger number of lower-mass stars with $M<1M_{\odot}$) than the Kroupa (\citealt{Kroupa_2001MNRAS_IMF}) or Chabrier (\citealt{Chabrier_2003ApJL_IMF}) IMFs that are observed for the Milky-Way Galaxy (e.g., \citealt{van_Dokkum_2010Natur_LowMassStarPopulation, Spiniello_2012ApJ_BotHeavyIMF, Conroy_2012ApJ_ETG_IMF_AbsorptionLine_II,La_Barbera_2013MNRAS_ETG_IMFfromSpecFeature,Spiniello_2014MNRAS_ETG_IMF}; but note that some studies favor Chabrier IMFs for massive early-type galaxies, e.g., \citealt{Smith_2015MNRAS_SINFONI,Sonnenfeld_2019A&A_SuGOHI_III}). Meanwhile, spatially varying IMFs inside massive early-type galaxies have also been revealed by data such that the very central region tends to have a more bottom-heavy IMF than at outer parts (e.g., \citealt{Martin-Navarro2015MNRAS_ETG_IMF, La_Barbera_2016MNRAS_RaidalConstraintsIMF, van_Dokkum_2017ApJ_ETG_IMF_AbsorptionLine_III, Parikh_2018MNRAS_MaNGA_IMF_spatial, Bernardi_2023MNRAS_MaNGA_IMF_systematic}). 

We note that accurate estimates on the IMF are essential in order to correctly predict the stellar mass through stellar population synthesis (SPS). The stellar mass can then be used to derive the dark matter mass, when further combined with dynamical and/or lensing measurements that provide estimates on the total mass (see e.g., \citealt{Newman_2017ApJ_SNELLS_IMF}). However, high-quality spectroscopic observations that can put stringent constraints on the IMF, via accurately identifying IMF-sensitive absorption features, are not cheaply available in particular for galaxies beyond the local Universe. The IMF is rather often taken as a basic assumption during SPS analyses. The inferred stellar mass may then be subject to a bias due to the difference between the referenced and the true IMFs. In this case, the true stellar mass can be derived through IMF-independent approaches. This is often done with the aid of stellar or gas kinematics, as well as gravitational lensing (e.g., \citealt{Cappellari_2013MNRAS_ATLAS_XX, Sonnenfeld_2018MNRAS_SL_WL,Shajib_2021MNRAS_NFW_DMHalo}). For example, one can choose to model the total density profile, either taking an observationally-motivated isothermal distribution or simply assuming ``mass follows light'' (e.g., \citealt{Koopmans_2006ApJ_SLACS_III, Smith_2015MNRAS_SINFONI}). Subtracting the best-modeled dark matter content from the total mass distribution then yields the stellar distribution. These gravitationally derived {\it stellar masses} are then treated as unbiased estimates, and then used to infer the true IMFs through comparisons to those obtained from the SPS approach. Up to now, studies in this regard found that the IMFs of spiral galaxies tend to have a normalization similar to a Chabrier/Kroupa IMF; assuming a Salpeter IMF normalization would simply results in a total mass that exceeds the one as required by kinematics or lensing measurements (\citealt{Bell_2001_ApJ_M2L-TFrelation,Kassin_2006_ApJ_BrightSprial-Star&DM,Brewer_2012MNRAS_SWELLS-III_LensedSpirals,Suyu_2012ApJ_LensedSpiral-B1933+503}). Conversely, studies of early-type galaxies typically found heavier Salpeter-like IMF (normalization) (e.g., \citealt{Treu_2010ApJ_ETGs_IMF, Auger_2010ApJ_LensIMF, Sonnenfeld_2012ApJ_ETG_SalpeterIMF, LiHongyu_2017ApJ_MaNGA_IMF_variation}), including those studies that use quasar micro-lensing observations to directly constrain a stellar mass fraction (e.g., \citealt{Oguri_2014MNRAS_Quasar_microlensing,Schechter_2014ApJ_FP_LensedQuasar,Jimenez-Vicente_2015ApJ_microlensing_fdm}).

In these lensing- and dynamics-based studies, to disentangle the contribution between dark matter and stars, the dark matter density distribution is often modeled either using a standard NFW profile (\citealt{NFW_1997ApJ}) or by a generalized NFW profile (gNFW, \citealt{ZhaoHongsheng_1996MNRAS_gNFW, Wyithe_2001ApJ_gNFW}), or via some generally contracted or cored halo density profiles (e.g., \citealt{Blumenthal_1986ApJ_DM_AC,Burkert_1995ApJ_DMhalo_DwarfGalaxy,Gnedin_2004ApJ_DM_ACmodel}). The latter cases allow for a variation in the inner slope of the dark matter halo, as a response to the central baryons (see \citealt{Cappellari_2013MNRAS_ATLAS_XV} for a variety of commonly assumed dark matter halo density profiles). On the stellar side, as the luminosity profile can be directly measured from spatially resolved images, the stellar mass distribution is often modeled by taking the best-fit luminosity profile and multiplying it with some assumed stellar mass-to-light ratio $M_{\star}/L$. 

The combined two-component model is then used to fit observational data. Here, we shall make a further note that the exact implementation for the above-mentioned modeling approaches is also subject to whether or not one has access to spatially resolved kinematic measurements. For nearby galaxies where integral field spectroscopic data are available, e.g., those from the Atlas$^{3D}$ (\citealt{Cappellari_2011MNRAS_ATLAS_I}) and the MaNGA (\citealt{Bundy_2015ApJ_MaNGA_Overview}) surveys, spatially-resolved mass reconstructions can be achieved through, for example, 
Jeans anisotropic modeling (JAM) methods (\citealt{Jeans_1922MNRAS, Cappellari_2008MNRAS_JAM, LiHongyu_2017ApJ_MaNGA_IMF_variation, Zhu2023_MaNGADynPopI, Lu2023_MaNGADynPopII}) or via particle- or orbit-based modeling techniques (\citealt{Syer_1996MNRAS_M2M, Schwarzschild_1979ApJ_SchwMod, ZhuLing18_CALIFA}). 

However, for galaxies at higher redshifts, in most cases the only available stellar kinematic measurements are single-aperture or long-slit velocity dispersions. Many studies have routinely combined such stellar kinematics with gravitational lensing, which put certain constraints on projected masses within some different radii. Specifically, the stellar velocity dispersion measured within a single aperture typically of a few kpc is a projected kinematic property of stars within the entire sight line through the stellar realm of a galaxy. Strong lensing phenomena often take place at a few kpc in projection from the galactic centre, the measurements can robustly provide a total (projected) mass normalization within the Einstein radius (see \citealt{Koopmans_2006ApJ_SLACS_III, Treu_2010ARA&A_GalStrongLensing}). While weak lensing phenomena take place at several hundreds of kpc, the measurements can put constraints on the total matter density profile at halo outskirts, where dark matter plays a dominant role (e.g., \citealt{Auger_2010ApJ_LensIMF, Schulz_2010MNRAS_SDSS_ETGs_WeakLensing, Sonnenfeld_2018MNRAS_SL_WL, Sonnenfeld_2019A&A_CMASS, Shajib_2021MNRAS_NFW_DMHalo}). Such jointly modeling approaches have been widely implemented for nearly a hundred strong lensing galaxies beyond the local Universe (see e.g., \citealt{Bolton_2012ApJ_BELLS_II, Treu_2010ApJ_ETGs_IMF, Sonnenfeld_2015ApJ_SL2S_V, Shu_2015ApJ_SLACS_XII, Oldham_2018MNRAS_SL, Shajib_2021MNRAS_NFW_DMHalo}). The majority of these lenses are obtained from the Sloan Lens ACS Survey (SLACS, \citealt{Bolton_2006ApJ_SLACS_I,Shu_2017ApJ_XIII}), and the Canada-France-Hawaii Telescope Legacy Survey (CFHTLS) Strong Lensing Legacy Survey (SL2S, \citealt{Cabanac_2007A&A_SL2S, Gavazzi_2012ApJ_SL2S_I}).   

Interestingly, studies in this regard so far have not reached consensus regarding the IMF normalization (and thus the central dark matter fraction and shape). The majority of the lensing galaxy population are relatively massive early-type galaxies (typically with stellar velocity dispersion of $\sigma \sim 200-350$ km/s). \citet{Treu_2010ApJ_ETGs_IMF} applied a joint model composed of two components to 56 SLACS lenses. The dark matter distribution therein was modeled by a standard NFW profile but with scale radius $r_{\rm s}$ fixed to 30 kpc. The light distribution was described by an Hernquist profile (\citealt{Hernquist_1990ApJ}). Under the assumption of a constant $M_{\star}/L$, a Salpeter IMF normalization was obtained. \citet{Auger_2010ApJ_LensIMF} adopted a slightly different modeling technique to the same galaxy sample. In their study, an Hernquist profile was directly assumed for the stellar {\it mass} density distribution, with the scale radius proportional to the V-band effective radius. A Salpeter-like IMF normalization was also obtained when assuming a standard NFW  halo. However, if a moderately contracted dark matter halo was adopted, the inferred IMF normalization became lighter than Salpeter but still heavier than Chabrier. \citet{Sonnenfeld_2015ApJ_SL2S_V} studied 81 early-type lensing galaxies from a combined sample of SL2S and SLACS. Through a hierarchical Bayesian modeling approach, population properties on central dark matter fraction and stellar IMF were inferred. Assuming a de Vaucouleurs light profile (\citealt{deVaucouleurs_1948AnAp}) with a constant $M_{\star}/L$ and a generalized NFW halo with scale radius $r_{\rm s}$ fixed to 10 effective radius $R_{\rm eff}$, the strong lensing plus stellar kinematics data preferred a Salpeter IMF normalization (while the dark matter inner slope could not be well constrained). \citet{Oldham_2018MNRAS_SL} studied the IMF normalization in 12 early-type lensing galaxies at $z=0.6$. Adopting a two-component model composed of a gNFW profile for the dark matter halo and a S{\'e}rsic profile (\citealt{Sersic_1963BAAA}) for the light distribution, the authors derived a Salpeter-like IMF normalization, as well as a clear bimodal distribution covering both cuspy and cored dark matter inner density slopes, under the constant $M_{\star}/L$ assumption.

However, some other studies obtained a different conclusion regarding the IMF normalization. For example, \citet{Sonnenfeld_2019A&A_CMASS} studied 23 massive galaxies from the Baryon Oscillation Spectroscopic Survey (BOSS, \citealt{Schlegel_2010_BOSS,Dawson_2013AJ_BOSS-III}) constant mass (CMASS) sample. The stellar mass was estimated by subtracting best-fit dark matter distribution (as constrained by weak lensing measurements at larger radii) from the total mass (as constrained by strong lensing at a few kpc). Then through a comparison to the SPS-inferred stellar mass, a normalization lighter than that of the Salpeter IMF was found within the strong lensing region. At more massive end ($\sigma > 300$ km/s), \citet{Smith_2013MNRAS_lightweight_IMF} and \citet{Smith_2015MNRAS_SINFONI} found Kroupa-like IMFs for three giant elliptical galaxies from the SINFONI Nearby Elliptical Lens Locator Survey (SNELLS). This was achieved through approximating the stellar mass by taking the difference between the lensing-derived total mass and a simulation-based dark-matter mass estimate (thus eliminating the necessity of making explicit assumptions on $M_{\star}/L$). Regarding the seemingly contradicting IMF trends between galaxies from SNELLS and SLACS/SL2S, \citet{Newman_2017ApJ_SNELLS_IMF} carried out a detailed analysis using several different methods on the same SNELLS galaxies in order to derive the IMFs therein. One of them is to combine strong lensing modeling with the axisymmetric JAM method to fit stellar kinematics. Again, a lighter IMF normalization was reached for the three SNELLS lenses, consistent with the previous conclusion from \citet{Smith_2013MNRAS_lightweight_IMF} and \citet{Smith_2015MNRAS_SINFONI}. If this is true, this would indicate that a wide scatter in the IMF normalization (thus star assembly history) may exist among massive early-type galaxies due to unknown reasons\footnote{Interestingly, while applying SPS modeling to high-quality spectroscopic data, \citet{Newman_2017ApJ_SNELLS_IMF} allowed for variations in the power law slope as well as a lower-mass cut off of the IMF. They found that in order to reach consistent results between lensing, dynamical and spectroscopic constraints, a power-law IMF cannot extend to the canonical hydrogen-burning limit of $0.08\,M_{\odot}$; otherwise the derived stellar mass would be in conflict (too large) with the total-mass bound from lensing constraints. A lower-mass cutoff of $m_{\rm cut}\sim 0.3\,M_{\odot}$ was proposed therein, which is roughly consistent with $m_{\rm cut} \ga 0.1\,M_{\odot}$ as derived by \citealt{Barnabe_2013MNRAS_LowMassCutOff_of_IMF} and \citealt{Spiniello_2015MNRAS_TotalDensity&LowMassEndSlope_of_IMF}. The uncertainty of the lower-mass cutoff simply further complicates the determination of galaxy IMFs.}.

One of the major uncertainties in these lensing- and dynamics-based studies lies in the assumption of constant $M_{\star}/L$.   
Mounting evidence has suggested the existence of a radial variation in $M_{\star}/L$ across galaxy populations (e.g., \citealt{Tortora_2011MNRAS_M-L_grad, Garcia-Benito_2019A&A_M-L_sptaial_CALIFA, GeJunqiang_2021MNRAS_MaNGA_M-L_color, Lu2023_MaNGADynPopII}). A radial gradient in $M_{\star}/L$ has also been gradually implemented in dynamical and lensing studies. For example, \citet{Oldham_2018MNRAS_M87_IMF} applied a Jeans analysis to multiple dynamical tracers in M87, a nearby massive brightest cluster galaxy in Virgo. The data strongly favors a radial $M_{\star}/L$ variation. As such, a Salpeter IMF was found to only dominate the central 0.5\,kpc, and a Chabrier IMF was needed in regions beyond. \citet{Sonnenfeld_2018MNRAS_SL_WL} combined the strong lensing and stellar kinematic measurements of 45 SLACS lenses, plus weak lensing measurements made by the Hyper-Suprime-Cam (HSC, \citealt{Aihara_2018PASJ_HSC}) survey for 1700 massive quiescent galaxies from the Sloan Digital Sky Survey (SDSS). The authors found that a radial gradient in $M_{\star}/L$ of 0.2 was needed, and the data favored a Chabrier IMF normalization plus a standard NFW dark halo. Interestingly, \citet{Shajib_2021MNRAS_NFW_DMHalo} took a similar approach, combining strong lensing, weak lensing and stellar kinematics measurements of 23 SLACS lenses. Adopting double-S{\'e}rsic profile to fit galaxy light and allowing for an anisotropic velocity profile as well as a radial variation in $M_{\star}/L$, they however, obtained a Salpeter-like IMF normalization together with a mildly expanded NFW halo. In addition, the radial slope in $M_{\star}/L$ as required by data was only as small as $0.02$, consistent with zero gradient in the mass-to-light ratio.

It is worth noting that even though pioneering studies (as mentioned above) have gradually moved on using power-law $M_{\star}/L$ models, a consensus towards the inference on the IMF and/or dark matter halo shape is still not in place. Meanwhile many dynamical studies that target modern IFU-kinematic galaxy surveys still take constant $M_{\star}/L$ as a basic assumption (e.g., \citealt{Zhu2023_MaNGADynPopI,  Zhu2023_MaNGADynPopIII, Lu2023_MaNGADynPopII}). The goal of this study is, using a realistic galaxy population from state-of-the-art cosmological simulations, to address following questions: (1) How well can a constant value of $M_{\star}/L$ be used to approximate the profile in early-type galaxies? (2) Assuming there are sufficient observables about the total matter distribution and the light distribution, how well can we constrain the average $M_{\star}/L$, as well as the central dark matter fraction $f_{\rm DM}$ and the inner dark matter slope $\gamma_{\rm DM}$, under the assumption of a constant $M_{\star}/L$? (3) How do the systematic biases differ among different models that are adopted to describe the dark matter and stellar distributions? (4) What are the reasons behind model failures in correctly predicting the above-mentioned dark matter and stellar properties under the assumption? 

To do so, we take an early-type galaxy sample from the cosmological hydrodynamic TNG100 simulation (\citealt{Springel_2018MNRAS_TNG_GalClustering,Nelson_2018MNRAS_TNG_GalColorBimodality,Pillepich_2018MNRAS_TNG_StellarMassCotent,Naiman_2018MNRAS_TNG_ChemicalEvo,Marinacci_2018MNRAS_TNG_RadioMagnet}). A two-component model fitting is carried out within a Bayesian inference framework. Specifically, we adopt a gNFW model to describe the dark matter density distribution. The model for stellar mass distribution is made to closely follow the light distribution (up to a constant $M_{\star}/L$ normalization factor), which is modeled using two different luminosity density profiles: a single-S{\'e}rsic profile and double-S{\'e}rsic profile (\citealt{Sersic_1963BAAA}).

We point out that except for galaxy matter and light profiles, we do not create any specific mock observations, such as galaxy spectra, multi-band photometries, or gravitationally lensed images. Instead we make the model to directly fit the {\it total} matter density distributions of simulated galaxies, which are not directly observable. However, by doing so, we can evaluate model degeneracy and bias that arise as a consequence of a constant $M_{\star}/L$ assumption, without being affected by the assumptions adopted in the dynamical and/or lensing modeling processes, such as an isotropic velocity dispersion distribution.

The rest of this work is organized as follows. In section~\ref{section:2}, we will introduce the simulation we used and the preliminary data selections(section~\ref{section:2.1}), profile descriptions (section~\ref{section:2.2}) and our two-component fitting models (section~\ref{section:2.3}). Secondly, in section~\ref{section:3}, we will present the investigation on the mass-to-light ratio radial variation in our simulated sample. In section~\ref{section:4.1}, we will present the results of using the two-component model to fit our mock data, the total density profile, without other constraints. After that, we explore the results of alternative models with constraints on the central mass-to-light ratio and dark matter component in the section~\ref{section:4.2}. The further discussions related to our results are provided in section~\ref{section:5}, including the limitation of numerical resolution (section~\ref{section:5.1}) and the stellar mass-to-light ratio model (section~\ref{section:5.2}). Finally, we will summarize our work in section~\ref{section:6}. In this work, we adopt the cosmology as \citealt{Planck_2016A&A_Cosmo} which is used in IllustrisTNG: $\Omega_{\rm M} = 0.3089$, $\Omega_{\Lambda} = 0.6911$ and $h = 0.6774$. 
 
\section{Methodology}
\label{section:2}
\subsection{Galaxy sample selection}
\label{section:2.1}

\begin{figure*}
    \centering
    \includegraphics[width=\textwidth]{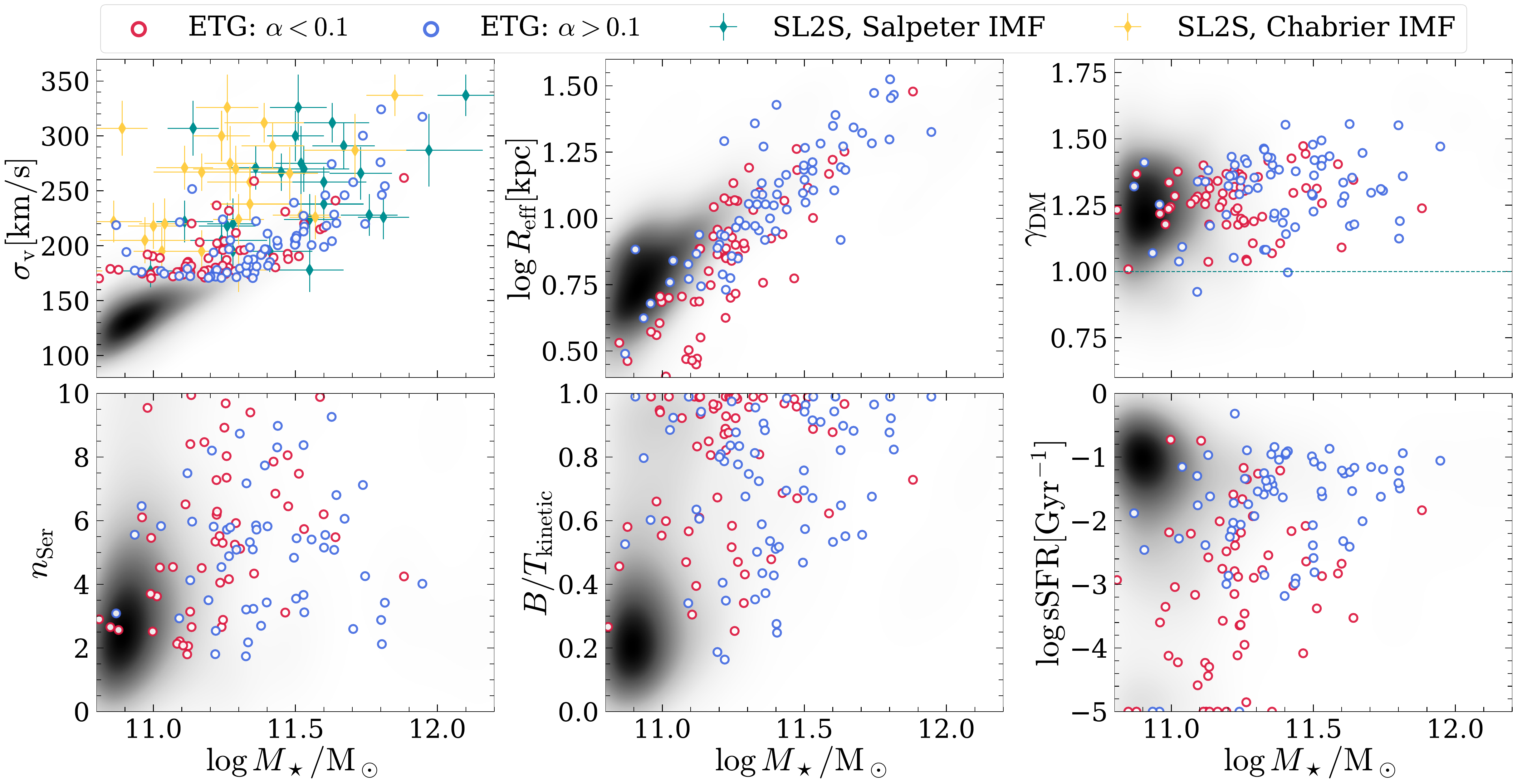}
    \caption{The fundamental properties as the function of total stellar mass of our sample from simulation. The red (blue) circles are our selected massive early-type galaxies with low (high) $M_{\star}/L$ radial gradient $\alpha$ (see section ~\ref{section:3} for detailed definition), while the shaded background represents the smoothed number density of remaining massive galaxies in the parameter space. In the top-left panel, we also present the stellar mass and velocity dispersion measurements of observational lensing galaxies in \citet{Sonnenfeld_2013ApJ_SL2S_III,Sonnenfeld_2013ApJ_SL2S_IV,Sonnenfeld_2015ApJ_SL2S_V} as diamonds whose colors indicate the different IMF assumption for stellar mass estimation. The horizontal dotted line in the top-right panel denotes the inner slope of the standard NFW model (\citealt{Navarro_1997MNRAS_SNoutflow-DMcore}). The bottom panels from left to right present, respectively, the S{\'e}rsic index, kinetic bulge-to-total ratio $B/T$, and the specific star formation rate within $2R_{\rm eff}$ (sSFR), as a function of the total stellar mass.
    }
\label{fig:sample_properties}
\end{figure*}

We utilize galaxies from The Next Generation Illustris Simulations (IllustrisTNG) (\citealt{Weinberger_2017MNRAS_AGNmod, Pillepich_2018MNRAS_IllustrisTNG, Pillepich_2018MNRAS_TNG_StellarMassCotent, Nelson_2018MNRAS_TNG_GalColorBimodality, Springel_2018MNRAS_TNG_GalClustering, Naiman_2018MNRAS_TNG_ChemicalEvo, Marinacci_2018MNRAS_TNG_RadioMagnet}). This is a suite of state-of-the-art magneto-hydrodynamic cosmological galaxy formation simulations, realized by the moving mesh code AREPO (\citealt{Springel_2010MNRAS_HydrodynamicsMovingMesh}). Galaxies in their dark matter halos are identified using the SUBFIND algorithm (\citealt{Springel_2001MNRAS_GalPopulation,Dolag_2009MNRAS_Substructure_Hydrodynamics}). For this work, we select galaxies from the TNG100 simulation, for which the simulation box size is $75/h\approx 110.7~\rm Mpc$; the mass resolution of the baryonic and dark matter are $m_{\rm baryon}=1.4\times 10^{6} {\rm M_{\odot}},m_{\rm DM}=7.5\times 10^{6} {\rm M_{\odot}}$, respectively; the gravitational soften length is $\epsilon = 0.74~\rm kpc$. The luminosity information of the simulated galaxies is associated with the star formation history. Each stellar particle is regarded as a stellar population containing stars born at the same redshift. The initial mass distribution of the stars in all stellar particles follows the Chabrier IMF (\citealt{Chabrier_2003ApJL_IMF}). During the evolution, the luminosity of the stellar particle is determined by the stellar population synthesis model GALAXEV (\citealt{Bruzual_2003MNRAS_SPS}). This particular simulation set has broadly reproduced many observed galaxy properties and scaling relations, for example, the mass-size relation (\citealt{Genel_2018MNRAS_TNG_SizeEvo}), the fundamental plane relation (\citealt{LuShengdong_2020MNRAS_TNG_FP}), galaxy density profiles (\citealt{WangYunchong_2019MNRAS_TNG_ETG_II,WangYunchong_2020MNRAS_TNG_ETG_I}), dark matter fractions (\citealt{Lovell_2018MNRAS_TNG_fDM}), as well as stellar orbit compositions (\citealt{XuDandan_2019MNRAS_TNGForb}). 

We remind the reader that the goal of this study is to demonstrate the consequence from a  constant $M_{\star}/L$ assumption that is commonly adopted in lensing and dynamical models. As a proof of concept, we select a galaxy sample that largely resembles that of SL2S. The SL2S galaxies are mainly early-type galaxies in a redshift range of $z=0.2\sim 0.8$. As shown in \citet{Sonnenfeld_2013ApJ_SL2S_IV}, their overall distributions in stellar mass, size and central velocity dispersion are similar to those of their lower-redshift counterparts from SLACS.   

In order to select a sample of simulated massive early-type galaxies that are representative of the observed lensing galaxies, we make use of the velocity dispersion, the total stellar mass, and the morphological properties as our criteria. First of all, we take central galaxies from the TNG100 simulation at $z = 0.5$ (snapshot 067), which is the average redshift of SL2S galaxies. We select galaxies according to their $r$-band luminosity-weighted central velocity dispersions $\sigma_{\rm v}$, which is measured along the $z$-projection (of the simulation box) within $0.5 R_{\rm eff}$ from the galaxy centre (where $R_{\rm eff}$ of simulated galaxy in this work is defined as the half-stellar-mass radius, i.e., half of galaxy stellar mass are distributed within the sphere of this radius, it is conceptually equivalent to the effective radius as defined in observations). The selected galaxies have $\sigma_{\rm v}$ spanning a similar range as the SL2S galaxies, i.e., satisfying $\sigma_{\rm v} \in [170,~350]\, {\rm km/s}$. This effectively corresponds to a stellar mass range of $\log{M_{\star}/{\rm M_{\odot}}} \in[10.8,~12.2]$. In addition, we implement the criteria similar to that adopted by \citet{XuDandan_2017MNRAS_IllustrisETG}. Here we recapture the key procedures therein. A two-component light model that integrates a de Vaucouleurs model (\citealt{deVaucouleurs_1948AnAp,Burkert_1993A&A}) to represent a bulge component, and an exponential model to describe a disk component, is used to fit the projected light distribution in the SDSS $r$-band of a galaxy. Such a bulge-disk decomposition provides an estimate of the photometric bulge-to-total ratio ($B/T$) as the quotient between the de Vaucouleurs luminosity and the total luminosity of the galaxy. The projected light distribution is also fitted individually using both de Vaucouleurs and exponential profiles. All model fittings to the projected light distributions are executed under elliptical symmetry, with ellipticity and position angle determined through luminosity-weighted second moment methods (for more details see section 2.2 of \citealt{XuDandan_2017MNRAS_IllustrisETG}). Here, we only select the galaxies (in their $z$-projection) with a photometric $B/T$ ratio exceeding 0.5, and further require that the de Vaucouleurs model provide a better fit compared to the exponential model for these galaxies (through comparing the $\chi^2$ of two individual light fits).

The top panels in Fig.~\ref{fig:sample_properties} show the velocity dispersion $\sigma_{\rm v}$ (left) and the effective radius $R_{\rm eff}$ (middle) as a function of the stellar mass $\log M_{\star}/M_{\odot}$ for the above-selected galaxies (red and blue circles), and for the remaining central galaxies within the same stellar mass range (as shaded background). As can be seen, there is a good agreement in the $\sigma_{\rm v}-M_{\star}$ distribution between the selected TNG100 galaxies and the SL2S sample (green and yellow diamonds). We can also see that the selected galaxies occupy the early-type sequence in the mass-size relation. The bottom panels present more morphological, dynamical and star-forming properties of the selected galaxies versus the remaining central galaxy population. It is evident that our further criteria essentially resulted in a sample dominated by early-type galaxies, the majority of which exhibit larger S{\'e}rsic index (>2), larger kinetic bulge-to-total ratio $(B/T)_{\rm kinetic}$, and lower specific star formation rate (sSFR). We note that the kinetic bulge-to-total ratio $(B/T)_{\rm kinetic}$ is taken from  \citet{XuDandan_2019MNRAS_TNGForb} and defined as twice the luminosity fraction of stellar particles with negative orbit circularities as used in \citet{Abadi_2003ApJ}. More fundamental properties of the simulated galaxies can be obtained from the TNG100 public catalogue (\citealt{Nelson_2015A&C_Illustris}). In the subsequent sections, we refer to the above-selected galaxies as the \textit{massive early-type galaxy} sample, including a total of 129 simulated galaxies. In section 3, we present their mass and light distributions, as well as radial profiles of $M_{\star}/L$.

In this study, we have taken a further step to form a sub-sample of massive early-type galaxies whose mass and light distributions are well approximated by models applied for this study (see section~\ref{section:2.2} for details), such that the final inferred biases would mainly stem from the assumption of a constant $M_{\star}/L$, instead of from inappropriate descriptions of the adopted models in the first place. To this end, we have first examined whether the dark matter 3D spherical profile of selected galaxies can be well approximated by the generalized NFW models (\citealt{ZhaoHongsheng_1996MNRAS_gNFW,Wyithe_2001ApJ_gNFW}), which will be adopted to describe dark matter in the two-component model. We find that gNFW profiles in general give a good description to the dark matter distribution of the simulated galaxies.

In comparison, the fitting performance using commonly adopted parametric models including the single-S{\'e}rsic, the double-S{\'e}rsic and the Hernquist model, to fit the projected light profiles of the simulated galaxies, in general has been less satisfactory. In sections~\ref{section:2.2} and~\ref{section:2.3}, we present a detailed description of the applied fitting procedure as well as our further criteria to classify a sub-sample of systems with ``good'' light-profile fitting. According to our criteria, among the 129 massive early-type galaxies, 41 (59) galaxies can be ``well'' fitted by the single- (double-) S{\'e}rsic model. It is noteworthy that, for each massive early-type galaxy, we employ our further criteria to determine whether it can be included into the sub-samples for both adopted light models. Thus these sub-samples are not mutually exclusive. In section~\ref{section:4}, we present the detailed model fitting results for this sub-sample of massive early-type galaxies in the TNG100 simulation. While if using an Hernquist model to fit the light profile, only a few galaxies would meet our criteria to be classified as a good fit. We do not therefore present detailed inference results for these few galaxies for the Hernquist light model. However, in the Appendix~\ref{section:B1}, we present a summary of model inference biases in $M_{\star}/L$ and central dark matter fraction $f_{\rm dm}$ for all 129 massive early-type galaxies in the sample. We must caution, however, the overall biases thereby would be a consequence due to both the assumption of a constant $M_{\star}/L$ and the adoption of inappropriate light models.
 
\subsection{Profile description}
\label{section:2.2}  
The light- and mass-profile fitting is a basic key ingredient in the following two-component model inference. As explained in the previous subsection, the projected light profile has been first directly fitted using a single- and double-S{\'e}rsic model (e.g., \citealt{Oldham_2018MNRAS_SL, Shajib_2021MNRAS_NFW_DMHalo}); the best-fit results have been used to assist a further sample selection, through which we fit composite models and estimate the biases coming from the assumption of constant $M_{\star}/L$. For these galaxies, the readily obtained best-fitted light models have then been de-projected and multiplied by a constant $M_{\star}/L$ factor in order to describe the $3D$ stellar mass profile. Together with a spherical gNFW profile describing the dark matter distribution, these two components are then combined to fit the total density profile data (see section~\ref{section:2.3}). The dark matter density profiles alone have also been fitted with a gNFW model. The best-fit model parameters obtained will provide a characteristic description of the dark matter distribution, serving as a ground truth reference to compare the two-component model inference results with.

To fit the mass and light profiles, we first bin the particle data (within a desired radial range) into spherical/circular shells according to their logarithmic distances from the centre of galaxy (the z-axis of the simulation box is regarded as the line-on-sight). We then fit the adopted model to the binned radial profile in the logarithmic space, assuming a relative error in density of 0.05 dex. Here below we give a detailed description of the models used. 


A few models have been developed to describe the density profile of a dark matter halo. Many works have demonstrated that different model assumptions will result in systematically different mass inferences in dynamical modeling (e.g.,\citealt{Scibelli_2019}). In this work, we fit a spherical gNFW profile (\citealt{ZhaoHongsheng_1996MNRAS_gNFW, Wyithe_2001ApJ_gNFW}) to the dark matter density distribution of a simulation galaxy within $R_{200}$, which is the radius that encloses an average total density of 200 times the critical density of the universe: 
\begin{equation}
\label{eq:gNFW}
    \rho_{\rm gNFW}(r) = \frac{\rho_{\rm DM}}{(\frac{r}{r_{\rm DM}})^{\gamma_{\rm DM}}(1+\frac{r}{r_{\rm DM}})^{3-\gamma_{\rm DM}}},
\end{equation}
where $\rho_{\rm DM}$, $r_{\rm DM}$ and $\gamma_{\rm DM}$, respectively, are the characteristic density, the scale radius and the inner density slope of the dark matter halo. We further define a concentration parameter $C_{\rm h}\equiv R_{200} / r_{\rm DM}$. Differences in the central steepness between the best-fit gNFW profile and the standard NFW profile (\citealt{NFW_1997ApJ}) are shown in top-right panel of Fig.~\ref{fig:sample_properties}. As can be seen, the simulated dark matter halos in general prefer steeper inner density profiles under the gNFW model, i.e., $\gamma_{\rm DM}>1$. We remark that such dark matter shape properties are consistent with recent lensing observations (e.g., see \citealt{Newman_2015ApJ_GroupScaleLenses}).


The profiles we use to fit the projected light distribution are the single- and double-S{\'e}rsic profile (\citealt{Sersic_1963BAAA,Baes_2019A&A_Sersic}):

\begin{equation}
\label{eq:Sersic}
    I_{\rm Ser}(R) = \frac{L b^{2m}}{2\pi m \Gamma(2m) R_{\star}^2}\exp{\left(-b(R/R_{\star})^{\frac{1}{m}}\right)},
\end{equation}
where $m$ is the S{\'e}rsic index and $L$ is the total luminosity of a given S{\'e}rsic component. The radius $R_{\star}$ is the projected half light radius of the individual S{\'e}rsic profile. The coefficient $b$ is defined through: $\gamma(2m,b) = \Gamma(2m)/2$ (where $\gamma(a,x)$ is the incomplete gamma function).

Once the best-fit single- or double-S{\'e}rsic model is obtained, the model $3D$ stellar mass density may come from the de-projection of the modeled light distribution multiplied with a constant projected $M_{\star}/L$. The analytic form of a de-projected S{\'e}rsic has been discussed in \citet{Baes_2011A&A_SersicAnalytic}, but in this work we adopt a numerical integration form for de-projected S{\'e}rsic model through a spherical Abel transform:
\begin{equation}
\label{eq:Abel}
    \rho_{\rm Ser}(r) = -\frac{\Upsilon_\star}{\pi}\int^{\infty}_{r}\frac{dR}{\sqrt{R^2-r^2}}\frac{d I_{\rm Ser}}{dR}.
\end{equation}

As we have mentioned in section~\ref{section:2.1}, we are in need of composing a sub-sample of galaxies, for which the light distribution shall be fairly well described by the adopted models in order to evaluate the true bias from constant $M_{\star}/L$ assumption. To do so, we have defined some fitting deviations $\Delta I_{\rm j}$ 
as follows:
\begin{equation}
    \Delta I_{\rm j} = \frac{1}{N_{\rm j}}\sum_{R_{\rm i}<R_{\rm ap,j}}|\log{I_{\rm i}}-\log{I_{\rm mod,i}}|
\label{eq:deviation}
\end{equation}
where $I_{\rm i}$ and $R_{\rm i}$ are the surface brightness and radius from galaxy centre of the $i_{\rm th}$ radial bin, respectively. And $I_{\rm mod,i}$ is the associated surface brightness from our light model (single- or double-S{\'e}rsic). The value $N_{\rm j}$ is the number of the radial bins within the $j_{\rm th}$ aperture with $R_{\rm ap,j}= (0.2R_{\rm eff},0.5R_{\rm eff},1R_{\rm eff},3R_{\rm eff})$. For our further sub-sample selection, we request galaxies to have all four $\Delta I_{\rm j} < 0.05 \,\rm dex$.

Furthermore, in the case where the effect of non-spherical symmetry becomes significant, the de-projected 3D light profile obtained through Eq.~(\ref{eq:Abel}) will also become less accurate in describing the true one and thus introduce additional biases. To eliminate so, we use the same criteria above, but replacing the projected surface brightness in Eq.~(\ref{eq:deviation}) with the $3D$ luminosity when evaluating $I_{\rm i}$, and the single- or double-S{\'e}rsic profile  with their de-projected models when evaluating $I_{\rm mod,i}$. We set the threshold in this case to be $\Delta I_{\rm j} < 0.1 \,\rm dex$. Such criteria above finally result in 41 (59) galaxies that can be ``well'' fitted by single- (double-) S{\'e}rsic models, which will be further discussed in section~\ref{section:4}. It is worth noting that the threshold for the de-projection case has been set at $<0.1 \rm dex$ to guarantee an adequate number of galaxies in the final sub-sample, simultaneously excluding galaxies that may potentially interfere with our analysis. To ensure the robustness of following results, we also conduct a test where we unify the thresholds for both projection and de-projection cases at $<0.075 \rm dex$, leading to smaller sub-sample: 30 (37) for the single-(double-)S{\'e}rsic light model scenario. In such a test, we have observed that the statistical results remain unchanged. Consequently, we conclude that the thresholds employed here are reasonable.

\subsection{Composite model fitting} 
\label{section:2.3}

\begin{table}
    \centering
    \caption{Ranges for the flat priors of model parameters}
    \begin{tabular}{|l|c|c|c|c|}
    \hline
    \hline
    & $\log{\Upsilon_\star/\Upsilon_\odot}$                     & $f_{\rm DM}$ & $\gamma_{\rm DM}^{\rm inner}$ & $\log{r_{\rm DM }}$        \\ \hline
    lower limit & \multicolumn{1}{l|}{$\log{\Upsilon_0/\Upsilon_\odot}-1 $} & 0.001        & 0.2                           & $\log{r_{\rm DM, true}}-1.2$ \\
    upper limit & $\log{\Upsilon_0/\Upsilon_\odot}+1 $                      & 0.999        & 1.8                           & $\log{r_{\rm DM, true}}+1.2$ \\ \hline
    \end{tabular}
    \label{tab:prior}
\end{table}

In this work, we have carried out the two-component model fitting procedure within a standard Bayesian framework through the implementation of the Markov Chain Monte Carlo (MCMC) method. The adopted procedure has been motivated by and closely follows the two-component jointly lensing-dynamical modeling methods used in observations (\citealt{Sonnenfeld_2015ApJ_SL2S_V,Sonnenfeld_2018MNRAS_SL_WL,Oldham_2018MNRAS_SL,Shajib_2021MNRAS_NFW_DMHalo}). In order to estimate modeling biases, we adopt the median of the posterior as the best-fit results and compare the results to the ground truth for the simulated galaxies. Here below we introduce the detailed data design, adopted models and parameter priors.

We compose two kinds of data for two-component model fitting. One is the total density profile within $4R_{\rm eff}$. These data are made to resemble situations where stellar kinematics and strong lensing measurements have been jointly used (e.g., \citealt{Sonnenfeld_2015ApJ_SL2S_V}). The other set is the total density profile within $R_{200}$, resembling situations where weak lensing observations at much larger (projected) radii are also available and used in addition (e.g., \citealt{Sonnenfeld_2018MNRAS_SL_WL,Shajib_2021MNRAS_NFW_DMHalo}).

We denote the model parameter vector as $\boldsymbol{\theta}$ and the profile data vector as $\boldsymbol{d}$. According to Bayesian theorem, we have
\begin{equation}
P(\boldsymbol{\theta}|\boldsymbol{d})\propto P(\boldsymbol{d}|\boldsymbol{\theta})P(\boldsymbol{\theta}),
\end{equation}
where $P(\boldsymbol{d}|\boldsymbol{\theta})$ is the data likelihood, $P(\boldsymbol{\theta})$ is the prior function depending on our experience or knowledge on model parameters, and $P(\boldsymbol{\theta}|\boldsymbol{d})$ is the posterior distribution that we sampled through MCMC. We assume a Gaussian error behavior to calculate the likelihood function: 
\begin{equation}
P(\boldsymbol{d}|\boldsymbol{\theta})\propto \exp \bigg(- \sum^{N}_{j=1}\frac{[\log\rho_{j}-\log\rho_{ j,\,\rm mod}(\boldsymbol{\theta})]^2}{2\epsilon_{\log\rho}^2}\bigg),
\end{equation}
where $\rho_{j}$ and $\rho_{j,\,\rm mod}(\boldsymbol{\theta})$ are respectively, the true mass density and the model predicted density given parameter $\boldsymbol{\theta}$ for the $j$th radial bin in logarithmic radius. We set the standard deviation $\epsilon_{\log \rho}$ for the logarithmic density to be 0.1 dex (see Appendix~\ref{section:A1}). We use a MCMC sampler via the {\sc emcee} python routine developed by \citet{Foreman-Mackey_2013PASP_EMCEE} to sample the parameters and generate the posterior distribution. 

The two-component model is made of two ingredients. The stellar mass density model is built based on light. We first find the model parameters that best fit the projected light distribution within $3R_{\rm eff}$ for a given light model investigated (i.e., either a single- or a double-S{\'e}rsic, see section~\ref{section:2.2}). This best-fit light profile is then de-projected to $3D$ and multiplied by a constant ratio $\Upsilon_\star\equiv M_{\star}/L$, in order to describe the stellar mass density profile\footnote{We point out that the ratio between the model prediction and the ground truth $\Upsilon_{\star,\rm mod}/\Upsilon_{\star,\rm true}$ essentially reflects the bias in stellar mass and thus the IMF normalization from dynamical or lensing models.}. The dark-matter density is modeled by a gNFW profile but with different arrangements in either pre-fixed or free parameters. In particular the normalization of the dark matter density profile is implicitly expressed via a $3D$ dark matter mass fraction $f_{\rm DM}$, which is defined as $f_{\rm DM}\equiv M_{\rm DM}(\leqslant R_{\rm eff})/[M_{\rm DM}(\leqslant R_{\rm eff}) +M_{\star}(\leqslant R_{\rm eff})]$. We note that both $\Upsilon_\star$ and $f_{\rm DM}$ are direct model parameters and shall be sought for through a Bayesian inference approach, which fits the composed model to the total density distribution. Here below we give a detailed summary of the four adopted models in a decreasing complexity order.

\subsubsection{An ``all-free'' model and all flat priors}
\label{section:2.3.1}
In this work, our baseline model consists in applying the two-component model to fit the total density profile directly and only considering the flat prior for all model parameters (note that the stellar mass profile is constrained by the known light distribution, only the mass-to-light ratio is the model parameter). Weak lensing measurements have been shown to effectively provide constraints on halo density profile at larger radii, where dark matter dominates over baryons. This can be used in combination with strong lensing and stellar kinematics at smaller radii in order to disentangle the contribution between the dark matter and baryons (e.g., \citealt{Sonnenfeld_2018MNRAS_SL_WL}). To achieve such constraints at larger radii, we take the total density profiles of galaxies up to $R_{200}$. For this data set, we adopt a full gNFW model to describe the dark-matter density distribution. This model is referred to as ``all-free'' and the parameters are $\boldsymbol{\theta} = \{ \log \Upsilon_\star/\Upsilon_\odot,\,f_{\rm DM},\, \gamma_{\rm DM},\,r_{\rm DM}\}$, where $\Upsilon_\odot \equiv \rm M_\odot/L_\odot$ is defined as the solar mass-to-light ratio. We adopted flat priors for all model parameters in this case. The upper and lower limits of the flat priors are given in Table~\ref{tab:prior}. In particular, the range for the logarithmic mass-to-light ratio $\log{\Upsilon_\star/\Upsilon_\odot}$ is set to be $[\log{\Upsilon_0}/\Upsilon_\odot-1,\log{\Upsilon_0/\Upsilon_\odot}+1]$, where $\Upsilon_0$ is the projected stellar mass-to-light ratio $M_{\star}/L$ measured within $R_{\rm eff}$ of the galaxy (i.e., the ratio of projected mass and light enclosed within $R_{\rm eff}$). In the subsequent discussions, this quantity also serves as the ground truth of the stellar mass-to-light ratio in the following discussions. Observationally this information can be obtained through SPS given a reference IMF, and therefore can be biased by a factor of $\sim 2-3$ due to an unknown IMF and other systematics.

\subsubsection{A gNFW dark-matter model and a Gaussian prior on $M_{\star}/L$}
\label{section:2.3.2}

The second case applied the same model as described in section~\ref{section:2.3.1}, with model parameters $\boldsymbol{\theta} = \{ \log \Upsilon_\star/\Upsilon_\odot,\,f_{\rm DM},\, \gamma_{\rm DM},\,r_{\rm DM}\}$. However, instead of a flat prior for $\log \Upsilon_\star/\Upsilon_\odot$, we adopt a Gaussian prior: 
\begin{equation}
\label{eq:m2l_prior}
    P(\log\Upsilon_\star/\Upsilon_\odot) \propto \exp{\left(-\frac{(\log\Upsilon_\star/\Upsilon_0)^2}{2\epsilon^2_{\log \Upsilon_\star}}\right)}.
\end{equation} 
We note that the standard deviation $\epsilon_{\log \Upsilon_\star}$ is set to be artificially small, of 0.05 dex ($\sim 10\%$), in comparison to currently best observational constraints available (see e.g., \citealt{Conroy_2012ApJ_ETG_IMF_AbsorptionLine_II, Spiniello_2014MNRAS_ETG_IMF,  van_Dokkum_2017ApJ_ETG_IMF_AbsorptionLine_III, Bernardi_2023MNRAS_MaNGA_IMF_systematic}). Same as in the previous case, the data set that is used to fit the model to is the total density profiles of galaxies up to $R_{200}$. We note that this practice will surely reduce the bias on $f_{\rm DM}$. The main goal of this model is to investigate remaining possible bias on the dark matter shape parameter $\gamma_{\rm DM}$, in an extreme case, where artificially accurate estimates on stellar mass is available. 

\begin{figure}
    \centering
    \includegraphics[width=\columnwidth]{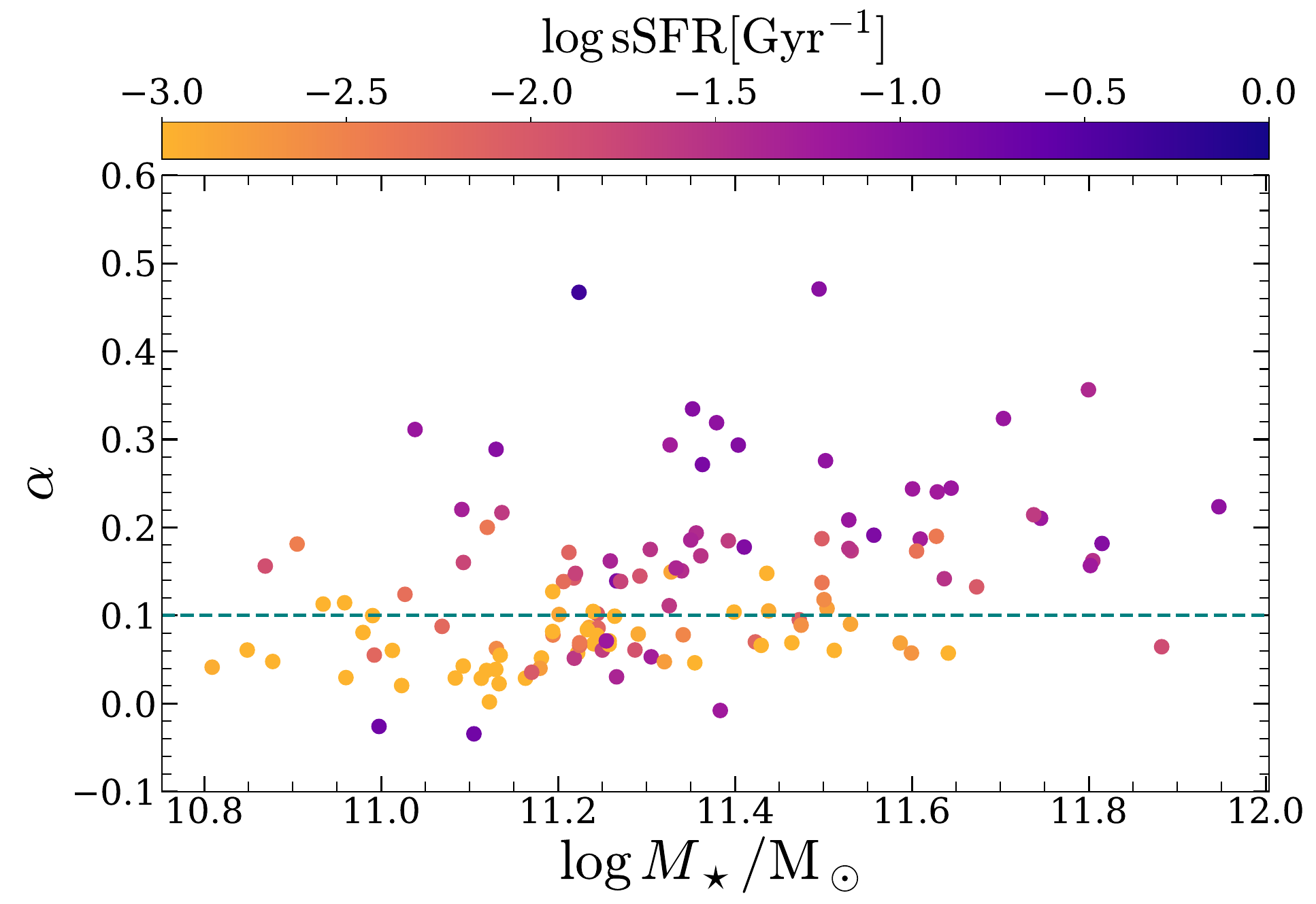}
    \caption{The stellar mass-to-light ratio gradient as the function of stellar mass for all 129 massive early-type galaxies in the sample. The $M_\star/L$ gradient $\alpha$ is defined as the power-law index of $\Upsilon_\star(R) = \Upsilon_\star(R_{\rm eff})(R/R_{\rm eff})^{-\alpha}$ fitted within $R_{\rm eff}$. All of the points are color-coded by the specific star formation rate within $2R_{\rm eff}$. The horizontal dashed line denotes the threshold of the constant/non-constant $M_\star/L$ galaxies.}
    \label{fig:m2l_alpha}
\end{figure}

\begin{figure}
    \centering
    \includegraphics[width=\columnwidth]{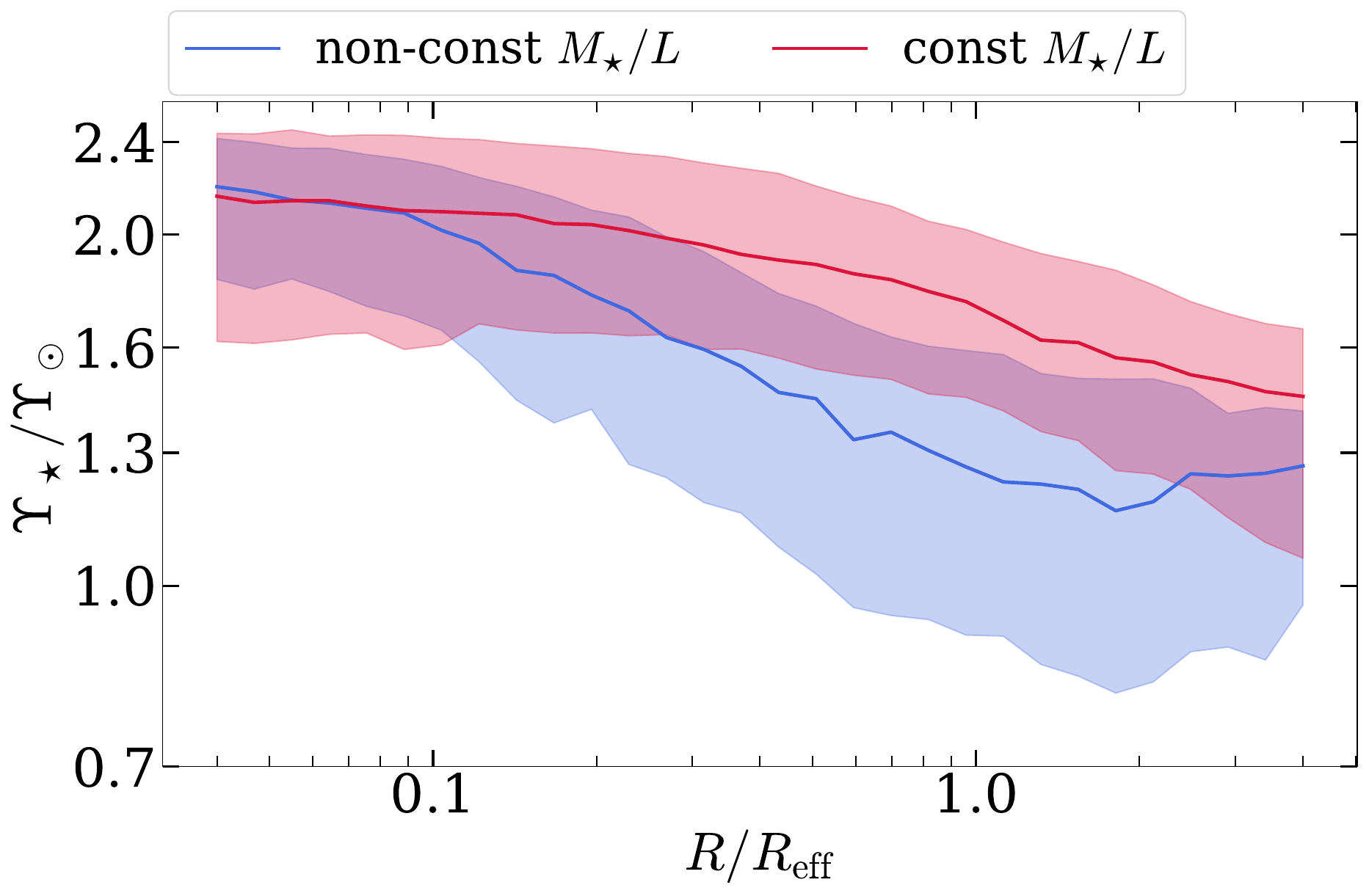}
    \caption{The stacked stellar mass-to-light radial profile of our massive early-type galaxy sample. We classify these galaxies into two categories (constant $M_\star/L$ systems: red, non-constant $M_\star/L$ systems: blue) according to their stellar mass-to-light ratio gradient. The solid lines are the median stellar mass-to-light ratio in radial bins of two categories. The shaded regions represent the $1\sigma$ scatter of the corresponding $M_\star/L$ radial profile. The unit of the y-axis is the solar mass-to-light ratio $\Upsilon_\odot\equiv \rm M_\odot/L_\odot$.}
    \label{fig:m2l_stack_prof}
\end{figure}

\subsubsection{An NFW dark-matter model and a flat prior on $M_{\star}/L$}
\label{section:2.3.3}

The third model aims at testing the cases where only stellar kinematic measurements (with or without strong lensing) are available, the spatial coverage of such data is often not large enough to constrain the overall shape of the dark matter distribution. Thus we limit the fit to $4R_{\rm eff}$. In addition, the gNFW model becomes a too flexible profile and we simplify it to a pure NFW by fixing the inner slope to $\gamma_{\rm DM}=1$ (\citealt{Treu_2010ApJ_ETGs_IMF}). In this model, the parameters are $\boldsymbol{\theta} = \{ \log \Upsilon_\star/\Upsilon_\odot,\,f_{\rm DM},\, r_{\rm DM}\}$. All of the priors $P(\boldsymbol{\theta})$ adopted are the same as described before.   

\begin{figure*}
    \centering
    \includegraphics[width=\textwidth]{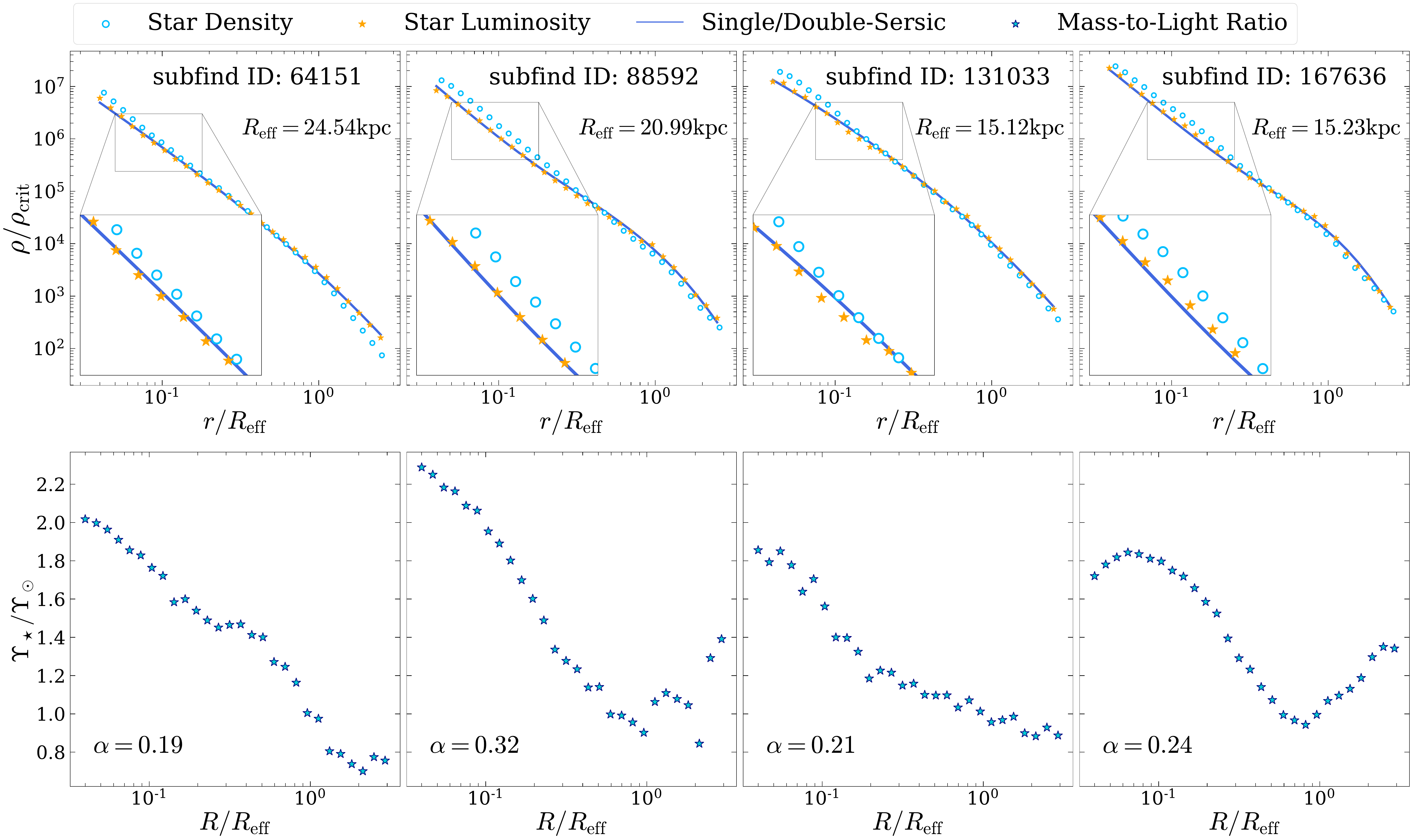}
    \caption{The de-projection of our light fitting and the stellar mass-to-light ratio profile. \textit{First Row:} The comparison of stellar mass density and the luminosity profile in the $3D$ space. Both the $3D$ luminosity profiles (yellow points) and our single/double-S{\'e}rsic models (blue line) are multiplied with the average stellar mass-to-light ratio within effective radius $\Upsilon_0$. The blue circles represent the real stellar density profile. \textit{Second Row:} The projected $2D$ stellar mass-to-light ratio radial profiles, where $\alpha$ is the power-law index of the stellar mass-to-light ratio measured within effective radius (see Eq.~\ref{eq:m2l_fitting}).}
    \label{fig:four_examples_light}
\end{figure*}

\subsubsection{A gNFW model with fixed $r_{\rm DM}$ and a flat prior on $M_{\star}/L$ }
\label{section:2.3.4}

An alternative modeling scheme to the one presented in section~\ref{section:2.3.3} is to fix the scale radius of the dark matter density, rather than fixing its inner slope. Motivated by \citet{Sonnenfeld_2015ApJ_SL2S_V}, we fixed the dark-matter scale radius $r_{\rm DM}$ to be $10R_{\rm eff}$. We also fitted this model to the total density profile data within $4R_{\rm eff}$. In this model, the parameters are $\boldsymbol{\theta} = \{ \log \Upsilon_\star/\Upsilon_\odot,\,f_{\rm DM},\, \gamma_{\rm DM} \}$. The priors are again all flat and the ranges are given in Table~\ref{tab:prior}.  In section~\ref{section:4}, we present the fitting results for all the adopted models described in this section.

\section{The stellar mass-to-light ratios of the simulated galaxies}
\label{section:3}
Before presenting the composite model fitting results and the systematic biases rooted in assuming a constant mass-to-light ratio $M_{\star}/L$ for the sub-sample of galaxies in section~\ref{section:4}, we first, in this section, display the actual spatial distribution of $M_{\star}/L$ for our full massive early-type galaxy sample. 

\begin{figure*}
    \includegraphics[width=\textwidth]{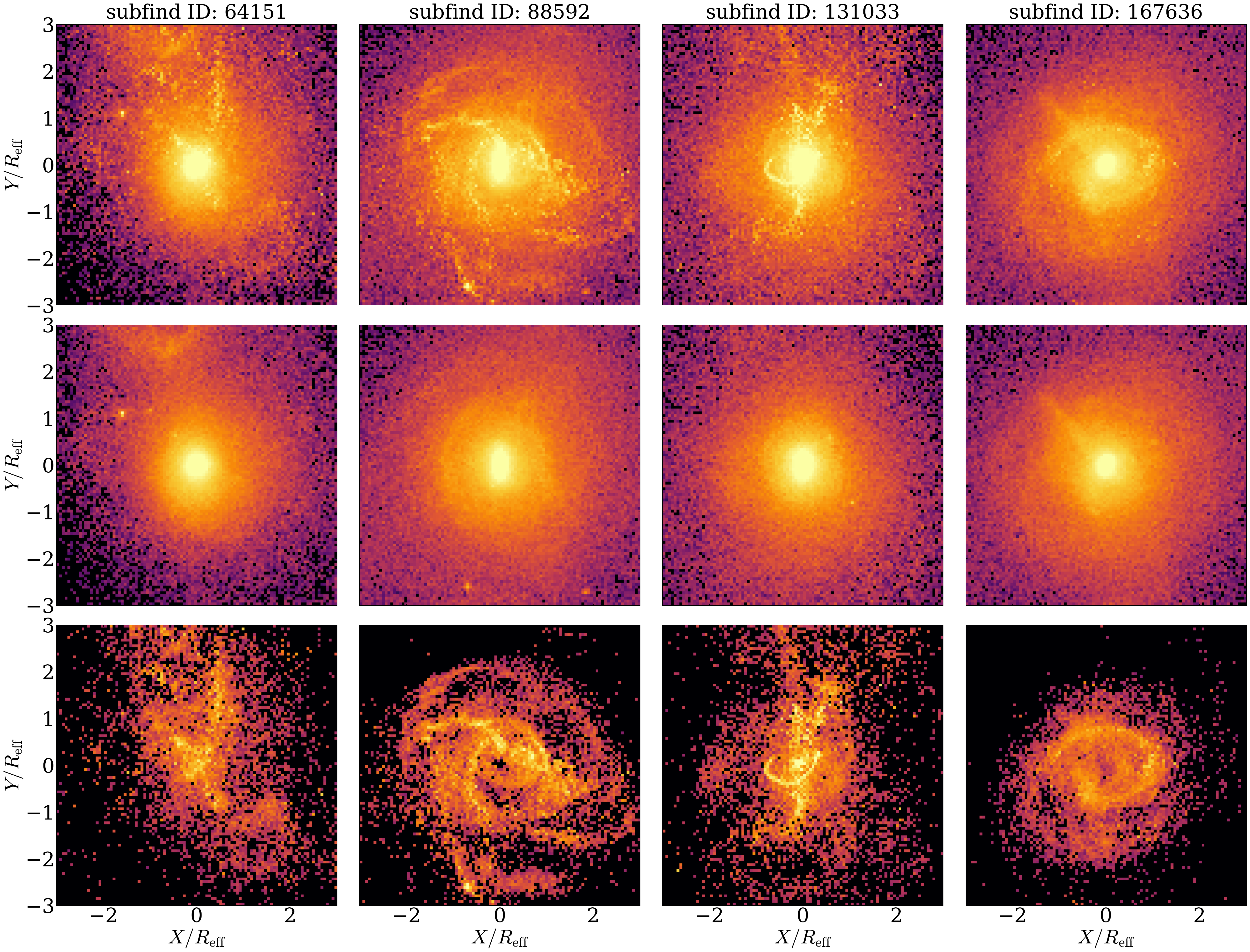}
    \caption{The face-on surface brightness map of the four examples in Fig.~\ref{fig:four_examples_light}. \textit{First Row:} The overall face-on surface brightness map of all stellar particles. \textit{Second Row:} The face-on surface brightness map of the stellar particles forming earlier than $1 \rm Gyr$. \textit{Third Row:} The face-on surface brightness map of the stellar particle with newly forming within $1 \rm Gyr$. The mass fractions for those particles forming within $1 \rm Gyr$ (shown in the bottom) are $3.84\%$, $4.73\%$, $5.40\%$, $4.47\%$ (from left to right). And their $M_\star/L$ gradients are 0.19, 0.32, 0.21, 0.24, respectively (see Fig.~\ref{fig:four_examples_light}).}
    \label{fig:four_examples_map}
\end{figure*}

In order to quantitatively describe the $M_\star/L$ radial variation, we adopt a simple power-law to fit the $M_\star/L$ profile within $R_{\rm eff}$:
\begin{equation}
\label{eq:m2l_fitting}
    \Upsilon_\star(R) = \Upsilon_\star(R_{\rm eff})\Big(\frac{R}{R_{\rm eff}}\Big)^{-\alpha}.
\end{equation}
where $\Upsilon_\star(R)$ represents the value stellar mass-to-light ratio around the given projected radius $R$. And the power-law index $\alpha$ is logarithmic gradient of $M_\star/L$. Without loss of generality, we simply divide our massive early-type galaxies into two categories. The galaxies with $\alpha$ exceeding 0.1 are classified as non-constant $M_\star/L$ systems while the remaining galaxies are still considered to exhibit a constant $M_\star/L$. Among our 129 massive early-type galaxies, there are totally 68 galaxies identified as a non-constant $M_\star/L$ system. In Fig.~\ref{fig:sample_properties}, such two categories are also labelled by blue and red, respectively. As can be seen already, the non-constant $M_\star/L$ systems are in general more massive and larger in size, as well as more actively star-forming, than their constant $M_\star/L$ counterparts. This can be seen again in Fig.~\ref{fig:m2l_alpha}, where the $M_\star/L$ gradient $\alpha$ is presented as the function of the total stellar mass and color-coded by the specific star formation rate within $2R_{\rm eff}$. We note that the distribution of $\alpha$ has a peak around $\sim 0.2$, which is consistent with observations, e.g., \citet{Sonnenfeld_2018MNRAS_SL_WL, Garcia-Benito_2019A&A_M-L_sptaial_CALIFA, GeJunqiang_2021MNRAS_MaNGA_M-L_color}. 

We present the stacked radial profiles of the two categories of galaxies in Fig.~\ref{fig:m2l_stack_prof}. As can be seen, constant $M_{\star}/L$ galaxies (red) on average have a flatter radial gradient than the non-constant $M_{\star}/L$ galaxies (blue). To provide a more comprehensive understanding of the properties of galaxies exhibiting non-constant $M_\star/L$, we have selected several representative examples to illustrate their characteristics.

The top panels of Fig.~\ref{fig:four_examples_light} provide the comparisons between the stellar mass density and luminosity profiles of four example non-constant $M_\star/L$ galaxies. The stellar mass (blue circles) and luminosity (yellow stars) density profiles are roughly compatible with each other around $R_{\rm eff}$, but the former is markedly more centrally concentrated than the latter (for the galaxies with stellar masses in a range as investigated in this study). This corresponds to generally increased $M_{\star}/L$ profiles towards galaxy centres, as can be seen from the bottom panels.

In Fig.~\ref{fig:four_examples_map}, we present the face-on projected brightness map of the same four galaxies. Notably, all four galaxies have star formation occurring during recent $1\rm\,Gyr$ out to $3R_{\rm eff}$ (see bottom panel). Despite the insignificant mass fraction of newly formed stars, they markedly decrease $M_{\star}/L$ at the outskirts. It is interesting to observe that there are galaxies (subfind ID: 88592, 167636) with a dented feature in the $M_{\star}/L$ profile as a consequence of a disk-like star formation in the past $1\rm\,Gyr$ within the region around $1-2R_{\rm eff}$.  

\begin{figure*}
    \centering
    \includegraphics[width=\textwidth]{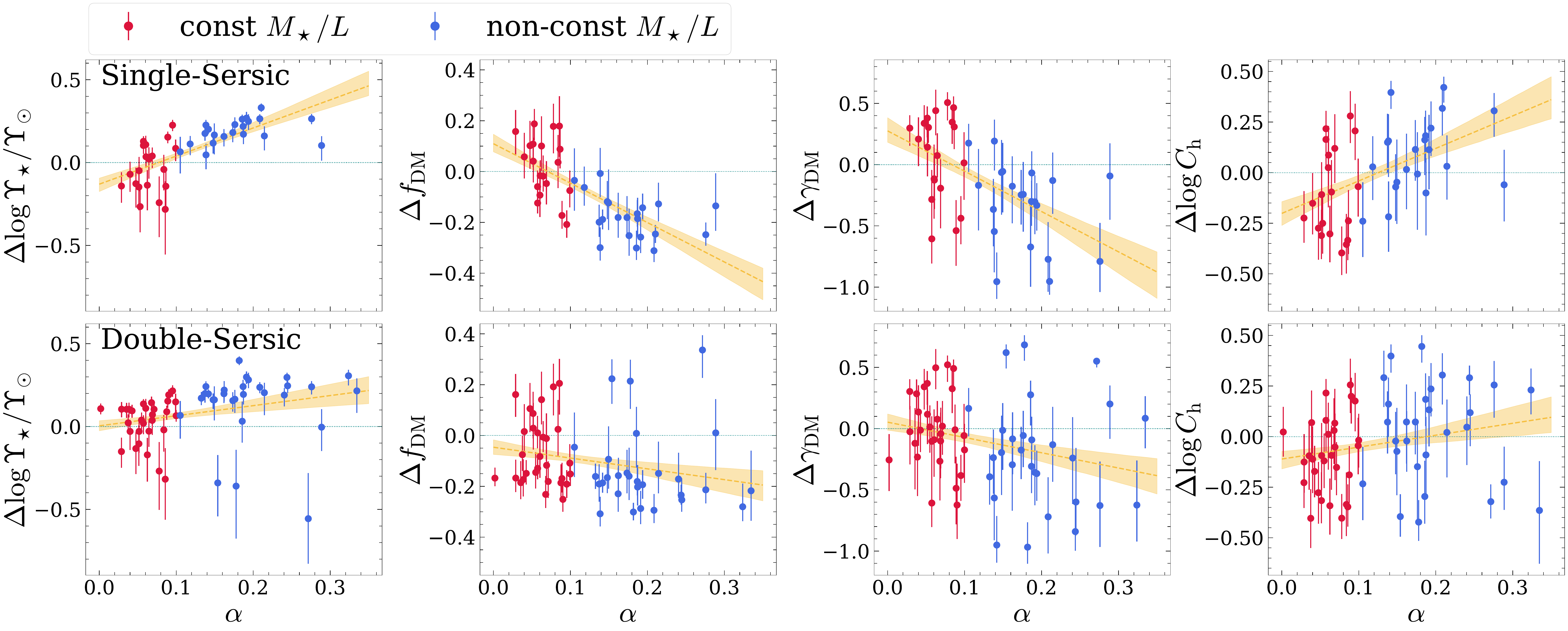}
    \caption{The biases between model predicted properties and the corresponding ground truth values, as a function of galaxy stellar mass-to-light ratio gradient. \textit{First Row:} single S{\'e}rsic light model (41 galaxies). \textit{Second Row:} double S{\'e}rsic light model (59 galaxies). The red points represent the systems with a constant stellar mass-to-light ratio ($\alpha<0.1$) while the blue points display the modeling bias of galaxies with a significant stellar mass-to-light ratio gradient ($\alpha>0.1$). The dashed dark yellow line (best-fit linear model) with the shaded area ($1\sigma$ error of bootstrap resampling) provide the trend between the modeling bias and the stellar mass-to-light ratio gradient. The detailed statistical biases are listed in Tab.~\ref{tab:para_bias_scatter}.}
    \label{fig:bias_alpha}
\end{figure*}

It is noteworthy that the mass-to-light ratio of the galaxy depends on many physical properties including the stellar IMF, the metallicity and the star forming history (as have been mentioned in the introduction). In observations, the radial gradient of the mass-to-light ratio is consistently recognized as an indicator of the radial variation in the stellar IMF (\citealt{Oldham_2018MNRAS_M87_IMF,Oldham_2018MNRAS_SL,Sonnenfeld_2018MNRAS_SL_WL}). In those works, researchers proposed that the stellar IMF might be heavier in the centre but keep the MW-like IMF in the outskirts, and such a centrally heavier IMF could exacerbate the declining trend of $M_\star/L$ profile. However, in current simulations, the luminosity of stellar particles is essentially generated under a uniform IMF (the Chabrier IMF in the TNG100 simulation), whereas the gradient in the mass-to-light ratio is primarily attributed to stellar particles that form in distinct epochs. As depicted in Fig.~\ref{fig:m2l_stack_prof}, the simulated $M_\star/L$ also exhibits a declining radial profile, which is similar with the $M_\star/L$ inferred by the centrally heavier IMF scenario in those previous works. Therefore, although the $M_\star/L$ variation in the real universe is attributed to more intricate factors, these simulated non-constant $M_\star/L$ galaxies can serve as a tool for discussing the potential biases that arise from the assumption of a constant $M_\star/L$. 

\begin{table*}
    \caption{The statistical biases with $1\sigma$ scatter of best-fit model parameters which are defined as the mean of the difference between the model predicted values and the ground truth. We also display the model predicted values and the ground truth of each modeling case (as well as their scatter). The bottom two rows show the linear slope of the trend between the biases and the stellar mass-to-light ratio gradient (see the dashed dark yellow line of Fig.~\ref{fig:bias_alpha}) while the Spearman's $\rho$(p-value) of such linear trends are presented in the bottom.}
    \centering
    \subtable[Light Model: Single-S{\'e}rsic]{
    \begin{tabular}{c|c|c|c|c|c}
         \hline
         \hline
         & & $\log\Upsilon_\star/\Upsilon_\odot$ & $ f_{\rm DM}$ & $ \gamma_{\rm DM}$ & $\log C_{\rm h}$ \\
         
         \hline
         const $M_\star/L$  & mean bias      & $-0.04\pm0.14$ & $0.02 \pm 0.12 $ & $0.08 \pm 0.34$ &$-0.10\pm 0.21$ \\
         &ground truth   &
         $0.26\pm0.05$  & $0.59 \pm 0.05 $ & $1.26 \pm 0.11$ &$0.74\pm 0.19$ \\
         &model predicted &
         $0.22\pm0.16$  & $0.61 \pm 0.11 $ & $1.33 \pm 0.32$ &$0.64\pm 0.17$ \\
         non-const $M_\star/L$  & mean bias  & $ 0.19\pm0.07$ & $-0.18\pm 0.08 $ & $-0.33 \pm 0.33$&$ 0.09\pm 0.18$\\
         &ground truth   &
         $0.19\pm0.08$  & $0.62 \pm 0.07 $ & $1.36 \pm 0.08$ &$0.60\pm 0.15$ \\
         &model predicted &
         $0.38\pm0.09$  & $0.44 \pm 0.10 $ & $1.03 \pm 0.34$ &$0.69\pm 0.20$ \\
         linear trend &$k_\alpha$                        & $1.70$         & $-1.55         $ & $-3.29$         & $1.61$\\
         
         &Spearman’s $\rho$                 & $0.78(<0.0001)$& $-0.76(<0.0001)$ & $-0.56(0.0001)$ &$0.50(0.0008)$\\
         \hline
    \end{tabular}
    }
    \subtable[Light Model: Double-S{\'e}rsic]{
    \begin{tabular}{c|c|c|c|c|c}
         \hline
         \hline
         & & $\log\Upsilon_\star/\Upsilon_\odot$ & $ f_{\rm DM}$ & $ \gamma_{\rm DM}$ & $\log C_{\rm h}$ \\
         \hline
         const $M_\star/L$  & mean bias      & $0.03 \pm 0.13$ & $-0.06\pm 0.13 $ & $0.00 \pm 0.30$ & $-0.09\pm 0.18$ \\
         &ground truth   &
         $0.30\pm0.06$ & $0.50 \pm 0.11 $ & $1.26 \pm 0.09$ &$0.81\pm 0.17$ \\
         &model predicted &
         $0.32\pm0.15$ & $0.44 \pm 0.20 $ & $1.26 \pm 0.29$ &$0.72\pm 0.16$ \\
         non-const $M_\star/L$ &mean bias  & $0.14 \pm 0.21$ & $-0.14\pm 0.16 $ & $-0.22 \pm 0.43 $ &$0.02\pm 0.24$\\
         &ground truth   &
         $0.16\pm0.09$ & $0.63 \pm 0.07 $ & $1.32 \pm 0.12$ &$0.63\pm 0.19$ \\
         &model predicted &
         $0.30\pm0.25$ & $0.49 \pm 0.15 $ & $1.11 \pm 0.41$ &$0.65\pm 0.22$ \\
         linear trend &$k_\alpha$                        & $0.61$          & $-0.43         $ & $-1.25$           &$0.59$\\
         
         &Spearman’s $\rho$                 & $0.55(<0.0001)$ & $-0.39(0.0023) $ & $-0.32(0.0137)$& $0.29(0.0262)$\\
         \hline
    \end{tabular}
    }
    \label{tab:para_bias_scatter}
\end{table*}

\section{Composite model fitting results} 
\label{section:4}

In this section, we report the inference results of the four composite models introduced in section~\ref{section:2.3}, and discuss reasons behind the statistical bias under the constant stellar mass-to-light ratio assumption. We primarily present our baseline model in section~\ref{section:4.1}, and summarize the results of three alternative models in section~\ref{section:4.2}.

\begin{figure*}
    \centering
    \includegraphics[width = \textwidth]{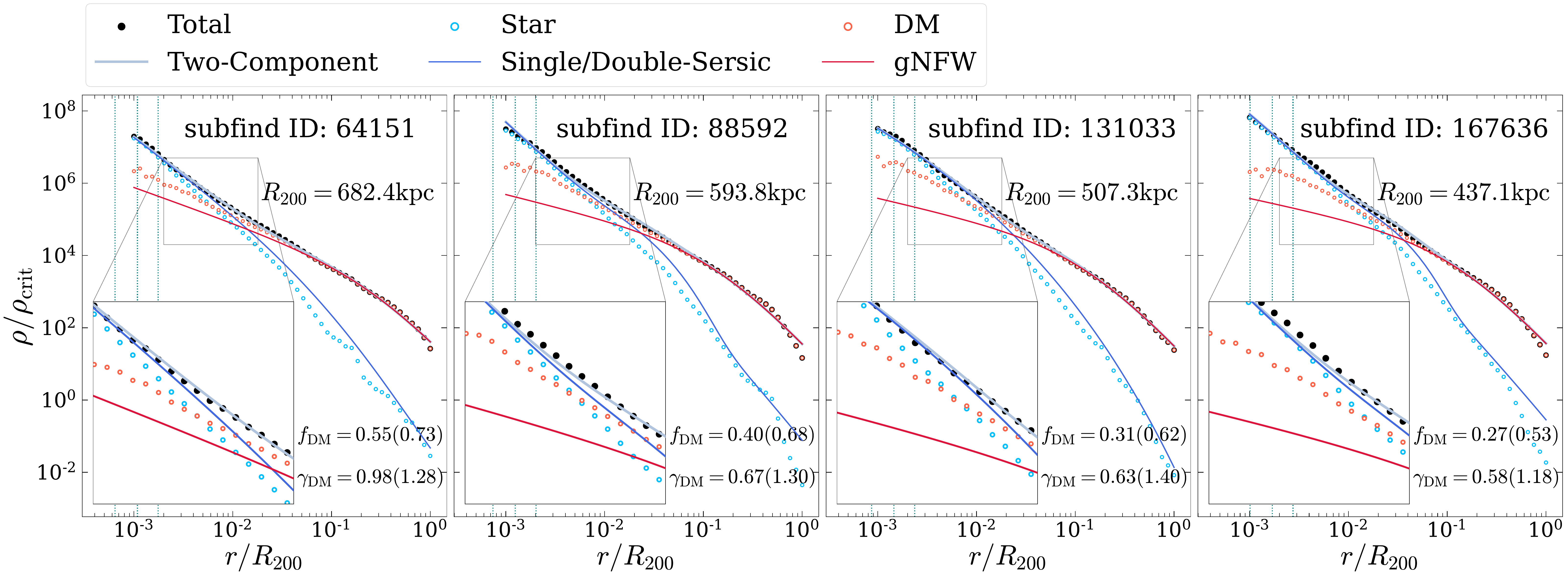}
    
    \caption{The examples of biased fitting results under constant stellar mass-to-light ratio assumption, containing the fitting results of our two-component density model (grey line), with the associated single/double-S{\'e}rsic model (blue line) for stars and the gNFW model (red line) for dark matter halo. The real density profiles are presented as points (black points: total density, blue circles: stellar density and red circles: dark matter density). The predicted dark matter fraction $f_{\rm DM}$ and inner slope $\gamma_{\rm DM}$ are listed in the bottom-right of the panel with the real values in the following brackets. The zoom-in panels illustrate the real stellar density slope is steeper than our single/double-S{\'e}rsic model while the latter is constrained by the light distribution.  Three dotted vertical lines in each panel indicate the region potentially suffering form the numerical resolution limitation. From left to right, these three vertical lines are the average radius of the artificial ``core'' ($0.44 \rm kpc$), the gravitational softening length of the simulation ($0.74 \rm kpc$) and the average radius within which the total density profile becomes shallower than the isothermal case ($1.2 \rm kpc$), respectively. More details about numerical resolution effect are provided in section~\ref{section:5.1}}
    \label{fig:four_examples_fitting}
\end{figure*}

\subsection{All-free model with a flat prior on mass-to-light ratio}
\label{section:4.1}
For this scenario, the model parameters are all free with flat priors and are constrained by the total density profile out to $R_{200}$. In Fig.~\ref{fig:bias_alpha} from left to right, we first show the statistical distributions of the modeling bias of each parameter versus the stellar mass-to-light ratio gradient $\alpha$, the former is defined as the difference between model predicted value and the value of the ground truth, the latter has been introduced in section~\ref{section:3}.

In this figure, the color of the points indicates the $M_\star/L$ category of each galaxy (also see section~\ref{section:3}). The top (bottom) panels display the results obtained using a single- (double-) S{\'e}rsic light model. As we have emphasized in section~\ref{section:2.2}, hereafter we only select the galaxies of which light distribution can be well described by our model, to further explore the biases that mainly arise from constant $M_\star/L$ assumption. It is clear that the non-constant $M_\star/L$ sample (blue) exhibits significantly more biased results. Specifically, the global stellar mass-to-light ratio $\log\Upsilon_\star/\Upsilon_\odot$ is overestimated in most non-constant $M_\star/L$ galaxies while the central dark matter fraction $f_{\rm DM}$ and dark matter inner slope are under-predicted by the best-fit model. The halo concentration has an opposite bias when compared to the dark matter inner slope. In general, there is a similar trend in the case of both light models that the bias increases with the stellar mass-to-light ratio gradient $\alpha$. We notice that, a few exceptions exist in the case double-S{\'e}rsic fitting, as their luminous (dark) components are under- (over-) predicted. We identify that these galaxies still suffer from both de-projection problem as well as the numerical resolution, which will be discussed in section~\ref{section:5.1}. As a reference, the galaxies with a constant $M_\star/L$ have a reduced bias, implying that our fitting method is robust and it is feasible to achieve unbiased inference for this type of galaxies.

In order to understand reasons behind such biases, we examined fitting results of individual galaxies. In Fig.~\ref{fig:four_examples_fitting}, we present the best-fit models against the true density profiles of four selected non-constant $M_\star/L$ galaxies in Fig.~\ref{fig:four_examples_light} . From left to right in roughly decreasing stellar-mass (virial radius) order. And the third galaxy is the result of considering a single-S{\'e}rsic model while the other three employ a double-S{\'e}rsic profile. Apparently, under-estimated dark matter fractions and maximal/enhanced central stellar mass densities (through over-estimated $\Upsilon_\star/\Upsilon_\odot$) are preferred in these systems. As already demonstrated in section \ref{section:3}, the stellar mass density can be more centrally concentrated than the light distribution. Through multiplying a constant $M_{\star}/L$, the de-projected light model would simply fail to capture such a central excess in true stellar mass density. In this case, the model has chosen over-predicting baryonic matter density (with the same profile as from the best-fit light distribution) to fit the total density directly (see the zoom-in panels of Fig.~\ref{fig:four_examples_fitting}). Therefore, an artificially maximal contribution from stars with over-predicted $\log\Upsilon_\star/\Upsilon_\odot$ are often preferred, also yielding under-prediction in $f_{\rm DM}$ and $\gamma_{\rm DM}$. 

The quantitative bias of each modeling parameter is listed in the Tab.~\ref{tab:para_bias_scatter}. Generally speaking, the non-constant $M_\star/L$ sample has a stellar mass-to-light ratio $\log\Upsilon_\star/\Upsilon_\odot$ over-estimated by $0.14$ -- $0.19\, \rm dex$ (depends on light model). As a consequence, the central dark matter fraction $f_{\rm dm}$ is under-estimated by $14\%$ -- $18 \%$. The model predicted inner slope of dark matter halo becomes shallower than the ground truth from fitting dark matter density profile alone, and is close to the standard NFW profile. Such biases could interfere our estimates and thus understanding of the stellar IMF estimation, dark matter steepness, as well as the interplay between baryons and dark matter in the galaxy centre.

\begin{figure*}
    \centering
    \includegraphics[width=\textwidth]{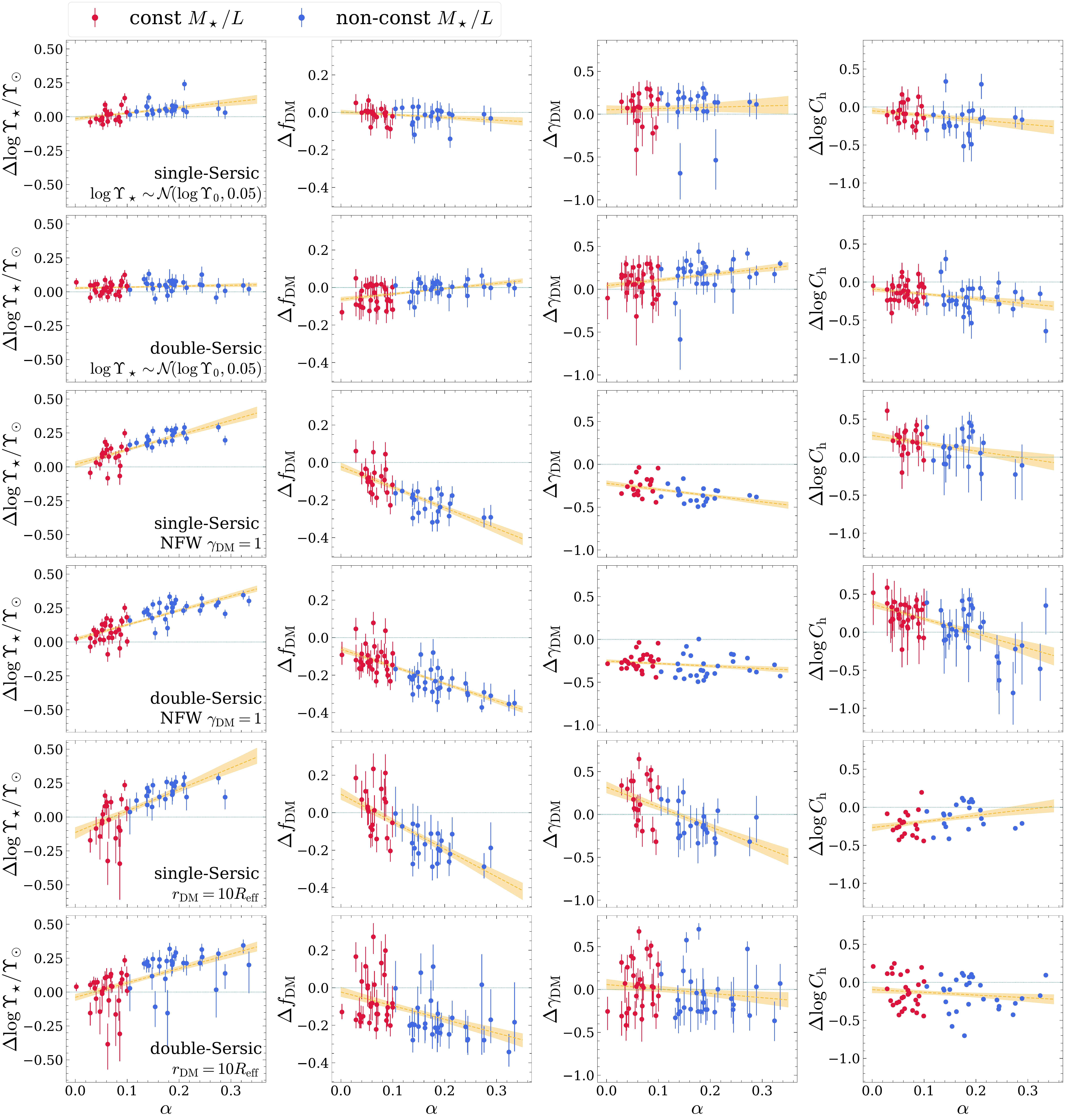}
    \caption{The modeling biases as the function of the mass-to-light ratio gradient of the three alternative models. Each column represents a specific model parameter whereas the rows are associated with the distinct scenarios (as indicated in the left panels). The color-coding scheme is quite similar with Fig.~\ref{fig:bias_alpha}. The shaded areas as well as the dashed dark yellow line still indicate the linear trends of the biases as the function of $\alpha$. We note that the error bar comes from the posterior sampling. Both the dark matter inner slope $\gamma_{\rm DM}$ in the NFW case and the halo concentration in the model of fixing $r_{\rm DM}$ at $10R_{\rm eff}$ are not real free modeling parameter. As a consequence, there is no error bar labelled on them. But their given values still have the differences from the ground truth which is presented as the bias.}
    \label{fig:results_alter}
\end{figure*}

\begin{table*}
    \caption{The similar table as Tab.~\ref{tab:para_bias_scatter} to present the modeling biases of three alternative models.}
    \centering
    \subtable[Fitting Assumption: $\log\Upsilon_\star\sim\mathcal{N}(\log\Upsilon_0,0.05)$,\,\, Light Model: Single-S{\'e}rsic]{
    \begin{tabular}{c|c|c|c|c|c}
         \hline
         \hline
         & & $\log\Upsilon_\star/\Upsilon_\odot$ & $ f_{\rm DM}$ & $ \gamma_{\rm DM}$ & $\log C_{\rm h}$ \\
         
         \hline
         const $M_\star/L$  & mean bias      & $0.01\pm0.05$ & $-0.01 \pm 0.04 $ & $0.05 \pm 0.18$ &$-0.07\pm 0.13$ \\
         &ground truth   &
         $0.26\pm0.05$  & $0.59 \pm 0.05 $ & $1.26 \pm 0.11$ &$0.74\pm 0.19$ \\
         &model predicted &
         $0.27\pm0.08$  & $0.58 \pm 0.09 $ & $1.31 \pm 0.18$ &$0.67\pm 0.15$ \\
         non-const $M_\star/L$  & mean bias  & $ 0.06\pm0.05$ & $-0.02\pm 0.04 $ & $0.08 \pm 0.24$&$ -0.17\pm 0.20$\\
         &ground truth   &
         $0.19\pm0.08$  & $0.62 \pm 0.07 $ & $1.36 \pm 0.08$ &$0.60\pm 0.15$ \\
         &model predicted &
         $0.25\pm0.09$  & $0.60 \pm 0.07 $ & $1.45 \pm 0.27$ &$0.43\pm 0.27$ \\
         \hline
    \end{tabular}
    }
    \subtable[Fitting Assumption: $\log\Upsilon_\star\sim\mathcal{N}(\log\Upsilon_0,0.05)$,\,\, Light Model: Double-S{\'e}rsic]{
    \begin{tabular}{c|c|c|c|c|c}
         \hline
         \hline
         & & $\log\Upsilon_\star/\Upsilon_\odot$ & $ f_{\rm DM}$ & $ \gamma_{\rm DM}$ & $\log C_{\rm h}$ \\
         \hline
         const $M_\star/L$  & mean bias      & $0.03\pm0.04$ & $-0.05 \pm 0.06 $ & $0.08 \pm 0.15$ &$-0.13\pm 0.12$ \\
         &ground truth   &
         $0.30\pm0.06$ & $0.50 \pm 0.11 $ & $1.26 \pm 0.09$ &$0.81\pm 0.17$ \\
         &model predicted &
         $0.32\pm0.08$ & $0.45 \pm 0.15 $ & $1.34 \pm 0.15$ &$0.68\pm 0.13$ \\
         non-const $M_\star/L$ &mean bias  & $0.04 \pm 0.04$ & $-0.01\pm 0.04 $ & $0.17 \pm 0.19 $ &$-0.22\pm 0.18$\\
         &ground truth   &
         $0.16\pm0.09$ & $0.63 \pm 0.07 $ & $1.32 \pm 0.12$ &$0.63\pm 0.19$ \\
         &model predicted &
         $0.21\pm0.11$ & $0.63 \pm 0.07 $ & $1.49 \pm 0.18$ &$0.41\pm 0.26$ \\
         \hline
    \end{tabular}
    }
    \subtable[Fitting Assumption: NFW $\gamma_{\rm DM} = 1$,\,\, Light Model: Single-S{\'e}rsic]{
    \begin{tabular}{c|c|c|c|c|c}
         \hline
         \hline
         & & $\log\Upsilon_\star/\Upsilon_\odot$ & $ f_{\rm DM}$ & $ \gamma_{\rm DM}$ & $\log C_{\rm h}$ \\
         
         \hline
         const $M_\star/L$  & mean bias      & $0.08\pm0.08$ & $-0.09 \pm 0.08 $ & $-0.26 \pm 0.11$ &$0.21\pm 0.17$ \\
         &ground truth   &
         $0.26\pm0.05$  & $0.59 \pm 0.05 $ & $1.26 \pm 0.11$ &$0.74\pm 0.19$ \\
         &model predicted &
         $0.34\pm0.09$  & $0.51 \pm 0.08 $ & $1.00 \pm 0.00$ &$0.95\pm 0.26$ \\
         non-const $M_\star/L$  & mean bias  & $ 0.22\pm0.05$ & $-0.23\pm 0.06 $ & $-0.36 \pm 0.08$&$ 0.11\pm 0.21$\\
         &ground truth   &
         $0.19\pm0.08$  & $0.62 \pm 0.07 $ & $1.36 \pm 0.08$ &$0.60\pm 0.15$ \\
         &model predicted &
         $0.41\pm0.08$  & $0.40 \pm 0.10 $ & $1.00 \pm 0.00$ &$0.71\pm 0.18$ \\
         \hline
    \end{tabular}
    }
    \subtable[Fitting Assumption: NFW $\gamma_{\rm DM} = 1$,\,\, Light Model: Double-S{\'e}rsic]{
    \begin{tabular}{c|c|c|c|c|c}
         \hline
         \hline
         & & $\log\Upsilon_\star/\Upsilon_\odot$ & $ f_{\rm DM}$ & $ \gamma_{\rm DM}$ & $\log C_{\rm h}$ \\
         \hline
         const $M_\star/L$  & mean bias      & $ 0.07\pm0.07$ & $-0.11\pm 0.07 $ & $-0.26 \pm 0.09$&$ 0.23\pm 0.17$ \\
         &ground truth   &
         $0.30\pm0.06$ & $0.50 \pm 0.11 $ & $1.26 \pm 0.09$ &$0.81\pm 0.17$ \\
         &model predicted &
         $0.37\pm0.08$ & $0.39 \pm 0.14 $ & $1.00 \pm 0.00$ &$1.04\pm 0.26$ \\
         non-const $M_\star/L$ &mean bias  & $0.24 \pm 0.06$ & $-0.25\pm 0.07 $ & $-0.32 \pm 0.12 $ &$0.00\pm 0.31$\\
         &ground truth   &
         $0.16\pm0.09$ & $0.63 \pm 0.07 $ & $1.32 \pm 0.12$ &$0.63\pm 0.19$ \\
         &model predicted &
         $0.41\pm0.10$ & $0.38 \pm 0.11 $ & $1.00 \pm 0.00$ &$0.63\pm 0.25$ \\
         \hline
    \end{tabular}
    }
    \subtable[Fitting Assumption: $r_{\rm DM} = 10R_{\rm eff}$,\,\, Light Model: Single-S{\'e}rsic]{
    \begin{tabular}{c|c|c|c|c|c}
         \hline
         \hline
         & & $\log\Upsilon_\star/\Upsilon_\odot$ & $ f_{\rm DM}$ & $ \gamma_{\rm DM}$ & $\log C_{\rm h}$ \\
         
         \hline
         const $M_\star/L$  & mean bias      & $-0.03\pm0.15$ & $0.01 \pm 0.12 $ & $0.18 \pm 0.26$ &$-0.23\pm 0.15$ \\
         &ground truth   &
         $0.26\pm0.05$  & $0.59 \pm 0.05 $ & $1.26 \pm 0.11$ &$0.74\pm 0.19$ \\
         &model predicted &
         $0.24\pm0.15$  & $0.60 \pm 0.11 $ & $1.44 \pm 0.20$ &$0.51\pm 0.08$ \\
         non-const $M_\star/L$  & mean bias  & $ 0.18\pm0.07$ & $-0.17\pm 0.08 $ & $-0.11 \pm 0.18$&$ -0.11\pm 0.16$\\
         &ground truth   &
         $0.19\pm0.08$  & $0.62 \pm 0.07 $ & $1.36 \pm 0.08$ &$0.60\pm 0.15$ \\
         &model predicted &
         $0.37\pm0.08$  & $0.45 \pm 0.10 $ & $1.25 \pm 0.15$ &$0.49\pm 0.08$ \\
         \hline
    \end{tabular}
    }
    \subtable[Fitting Assumption: $r_{\rm DM} = 10R_{\rm eff}$,\,\, Light Model: Double-S{\'e}rsic]{
    \begin{tabular}{c|c|c|c|c|c}
         \hline
         \hline
         & & $\log\Upsilon_\star/\Upsilon_\odot$ & $ f_{\rm DM}$ & $ \gamma_{\rm DM}$ & $\log C_{\rm h}$ \\
         \hline
         const $M_\star/L$  & mean bias      & $0.01\pm0.13$ & $-0.06 \pm 0.13 $ & $0.04 \pm 0.29$ &$-0.13\pm 0.21$ \\
         &ground truth   &
         $0.30\pm0.06$ & $0.50 \pm 0.11 $ & $1.26 \pm 0.09$ &$0.81\pm 0.17$ \\
         &model predicted &
         $0.31\pm0.15$ & $0.44 \pm 0.20 $ & $1.30 \pm 0.24$ &$0.68\pm 0.22$ \\
         non-const $M_\star/L$ &mean bias  & $0.19 \pm 0.11$ & $-0.18\pm 0.11 $ & $-0.06 \pm 0.27 $ &$-0.16\pm 0.21$\\
         &ground truth   &
         $0.16\pm0.09$ & $0.63 \pm 0.07 $ & $1.32 \pm 0.12$ &$0.63\pm 0.19$ \\
         &model predicted &
         $0.35\pm0.15$ & $0.46 \pm 0.11 $ & $1.26 \pm 0.21$ &$0.47\pm 0.08$ \\
         \hline
    \end{tabular}
    }
    
    \label{tab:para_bias_scatter_alter}
\end{table*}

\subsection{Alternative models}
\label{section:4.2}
\subsubsection{All-free model with a Gaussian prior on mass-to-light ratio}
\label{section:4.2.1}
In this subsection, we will explore whether a prior constraint on global $M_{\star}/L$ can help to break the model degeneracies and biased inference encountered in section~\ref{section:4.1}. Specifically, we present the results of an all-free model with a Gaussian prior on $\log \Upsilon_\star/\Upsilon_\odot$, centred on the true mass-to-light ratio $\Upsilon_0$ measured within $R_{\rm eff}$ and a standard deviation of $\sim 10\%$ (see section~\ref{section:2.3.2} for details).

The first two rows of Fig.~\ref{fig:results_alter} show the biases of key model parameters as the function of $M_\star/L$ gradient. In comparison to the ``all-free'' model (see Fig.~\ref{fig:bias_alpha}), the prediction on $f_{\rm DM}$ in general is much better, as expected, when $\log \Upsilon_\star/\Upsilon_\odot$ is provided with a Gaussian prior centred on the true value. However, it is interesting to notice a systematic overestimation on the dark matter inner slope $\gamma_{\rm DM}$ of those non-constant $M_\star/L$ galaxies, especially under double-S{\'e}rsic light model. That means the overestimation comes from the fact that the central density deficit between the true stellar mass density and the light-based model prediction can only be compensated for by an extra contribution from dark matter inner slope, as now we have ``fixed'' the level of baryonic matter through adopting a Gaussian $\log \Upsilon_\star/\Upsilon_\odot$ centred around the truth. In the first two subtables of Tab.~\ref{tab:para_bias_scatter_alter}, we list the mean bias of galaxies within individual $M_\star/L$ category in this modeling scenario.

This investigation demonstrates that, even if we could obtain unbiased measurement on stellar mass-to-light ratio within a given aperture (available at a single point), the constant $M_{\star}/L$ assumption could still be dangerous in causing artificially and significantly steepened innermost dark matter density distributions for galaxies in this mass range.

\subsubsection{An NFW dark-matter model and all flat priors}
\label{section:4.2.2}
For the model that adopts an NFW profile to fit the dark matter (see section~\ref{section:2.3.3}), the results are reported in the two middle rows of Fig.~\ref{fig:results_alter} and Tab.~\ref{tab:para_bias_scatter_alter}. The results of the two light model cases are highly consistent with each other. As can be seen, in both light model fitting cases (single- and double-S{\'e}rsic), and also regardless of galaxies having constant or non-constant $M_\star/L$, the dark matter fractions $f_{\rm DM}$ are now systematically under-estimated by roughly $10\%$ -- $25\%$, whereas correspondingly the mass-to-light ratios $\Upsilon_\star$ are over-estimated by $0.07\rm dex$ -- $0.24 \rm dex$ (depending on the light model, see Tab.~\ref{tab:para_bias_scatter_alter}). With the investigation as presented in previous cases, we can now readily understand such bias: as galaxies have inner dark matter slopes much steeper than the NFW prediction (see the third panel of Fig.~\ref{fig:sample_properties}), artificially fixing the inner dark matter slope (to be $1$ or other shallower values) would naturally result in underestimated (overestimated) dark matter fractions (mass-to-light ratios). 

\subsubsection{gNFW model with fixed scale radius }
\label{section:4.2.3}
In this subsection, we present the model results assuming the dark matter scale radius is fixed to be $r_{\rm DM} = 10 R_{\rm eff}$. Since the fitting range is confined within $4R_{\rm eff}$, the dark matter model can be approximately regarded as a simple power-law in the centre. Similar assumptions of fixing scale radius or halo concentration in dark matter model were commonly made in the previous works (\citealt{Koopmans_2003ApJ_LensETG_0047-281,Koopmans_2006ApJ_SLACS_III,Sonnenfeld_2015ApJ_SL2S_V}) when we cannot obtain the measurements to constrain the matter distribution of galaxy outskirts. Thus, this model can help us to investigate biases only with central constraints.  The statistical biases of this case are presented in two bottom rows of Fig.~\ref{fig:results_alter}. The dark matter fraction of non-constant $M_\star/L$ systems is comparatively under-estimated while the stellar constituent is enhanced.  

As can be seen from Tab.~\ref{tab:para_bias_scatter_alter}, for non-constant $M_\star/L$ galaxies, the model predicted stellar mass-to-light ratio $\log\Upsilon_\star/\Upsilon_\odot$ on average exceeds the ground truth by 0.18--0.19 $\rm dex$, while the model under-estimates the dark matter fraction by approximately $17\%$--$18\%$. Recent studies on central stellar kinematics and population for the nearby galaxies (\citealt{Zhu2023_MaNGADynPopIII}) indicates that neglecting the radial variation in the $M_\star/L$ also  leads to a generally lower dark matter fraction in the central regions than the results of accounting for the $M_\star/L$ gradient, which is roughly consistent with our analysis.

\section{Discussions}
\label{section:5}
\subsection{Numerical limitation}
\label{section:5.1}
In $N$-body simulations, gravitational softening is a necessary implementation in order to avoid close encounters of  particles and divergence of gravitational force in the collisionless regime. However, this can also introduce artificially smoothed structures of the matter distribution and the kinematics of galaxy below some typical softening length. Specifically, the density profile will have an artificial ``core'' in its innermost region due to such a numerical resolution problem, which may be beyond the fitting capability of our model. In TNG100, the gravitational softening length is $\sim 0.7 \,\rm kpc$. Actually, the region that suffers from the resolution issue is not only within this softening length. One of the reasons is that there is the energy re-partition between the baryon particle and dark matter particle originating from their different particle mass (\citealt{Ludlow_2019MNRAS_EnergyEquipartition,Ludlow_2023MNRAS_SpuriousHeating}).

In principle, the best way to explore the influence of numerical resolution is to take some simulation with a higher mass/force resolution and create a suite of runs with distinct parameters served as a convergence test (\citealt{Ludlow_2019MNRAS_ConvergenceTest}), which is beyond the focus of this work. Here we implement a more straight approach to assess the scale of the artificial ``core'' due to the resolution. We first extract the total density profile from $0.1\rm \,kpc$ to $100 \rm \,kpc$. Since the total density profile of the massive early-type galaxy is known to be approximately isothermal with a logarithmic radial slope close to 2, we apply softened power law to fit the total density profile to measure the scale of the artificial ``core'' which is significantly shallower than the isothermal model:
\begin{equation}
    \rho(r) \propto \frac{r_{\rm c}^{\gamma_{\rm tot}}}{(r^2+r_{\rm c}^2)^{\gamma_{\rm tot}/2}}
\end{equation}
where $r_{\rm c}$ is the core radius, and $\gamma_{\rm tot}$ is the radial slope beyond the innermost core. Upon the fitting result, we find that the isothermal slope can capture the total density distribution of our galaxies with $\gamma_{\rm tot} = 2.06 \pm 0.09$ beyond the innermost artificial ``core'' (see also \citealt{WangYunchong_2020MNRAS_TNG_ETG_I, WangYunchong_2019MNRAS_TNG_ETG_II}). And the core radius $r_{\rm c}$ of our massive early-type sample on average is $0.44 \pm 0.11 \rm kpc$. Between the general isothermal slope and the artificial ``core'', the averaged total density profile becomes shallower than the isothermal case within $\sim 1.2\rm\, kpc$.

We further investigate the effects of the numerical resolution in our modeling. As have been demonstrated, taking single-/double-S{\'e}rsic models to fit the light, the adoption of a constant stellar mass-to-light ratio can make the model under-estimate the dark matter and over-estimate the stellar component in the centre. On the other hand, we have also observed that both single-/double-S{\'e}rsic models are in fact not capable of fitting the innermost artificial ``core'' for the smallest galaxies in our sample. For these cases,  the model will then not have the freedom to artificially enhance the ``modeled'' stellar profile to fit the total density distribution at the centre, but only to enhance the central dark matter fraction and the dark matter inner slope to do so. 
This is clearly demonstrated by the several outliers in the bottom panel of Fig.~\ref{fig:bias_alpha}, where $\log\Upsilon_\star/\Upsilon_\odot$ ($f_{\rm dm}$) are extremely under-estimated (over-estimated). 
In Fig.~\ref{fig:four_examples_fitting}, we also use the vertical dotted lines to indicate both the softening length of TNG100 and the average scales of the innermost artificial ``core'' (i.e., $0.44\,\rm kpc$ as a core radius, $1.2\, \rm kpc$ indicating the radius where the central profile becomes significantly shallower than the isothermal case). 
We verify that for the majority of galaxies in our sample, the radial range where the central stellar mass distribution becomes significantly steeper than the central light distribution already exist beyond the central region that is potentially affected by the numerical resolution. 
We thus believe that our general conclusion will not be qualitatively interfered and thus changed due to the force softening effect.    

\subsection{The further discussion on stellar mass-to-light ratio model}
\label{section:5.2}
In section~\ref{section:3}, we have discussed the stellar mass-to-light ratio of our simulated galaxy. And the results in the  section~\ref{section:4} show that such a radial variation of $M_\star/L$ could lead to additional systematic biases of model inference under constant $M_\star/L$ assumption. In current research, several studies have considered a non-constant $M_\star/L$ model. For instance, a radial power-law  model has been adopted as an attempt to characterize the variation of stellar mass-to-light ratio (\citealt{Sonnenfeld_2018MNRAS_SL_WL,Oldham_2018MNRAS_M87_IMF,Oldham_2018MNRAS_SL,Shajib_2021MNRAS_NFW_DMHalo}). In this study, we have used the power law model to quantitatively measure the $M_\star/L$ gradient. We notice that, however, a power law model is not always necessarily a good approximation to describe the radial profile of $M_\star/L$ for the non-constant $M_\star/L$  galaxies in our sample (see the bottom panel of  Fig.~\ref{fig:four_examples_light}). Besides, adopting a power law model will simply add a new degree of freedom ($M_\star/L$ gradient $\alpha$), which will be degenerate with the previous parameters without additional constraints. It is suggested that the power law $M_{\star}/L$ model, to some degree, can inform us about the general steepness of the stellar mass-to-light ratio variation, but might not be a sufficient model from a dynamical modeling perspective, especially for those ``dented'' non-constant $M_\star/L$ systems. Another possible working model is to assign distinct $M_\star/L $ values to individual stellar components like disk and bulge (\citealt{Rigamonti_2022MNRAS_BD_m2l}). However, this model is more appropriate for the S0 galaxy whose disk component is more regular and massive than the cases of early-type galaxy in this work. In order to solve this problem, from the observation side, it would be helpful to find the additional constraints on the $M_\star/L$ variation from the spatially resolved color information or even spectroscopic data when we consider models beyond constant stellar mass-to-light ratio. From the simulation side, the application of a uniform IMF inherently limits our ability to identify the influence of IMF variation on the $M_\star/L$ distribution. Therefore, it is worthwhile to introduce the diversity of the IMF among stellar particles according to their formation environments as indicated by previous studies. (\citealt{Padoan_2002ApJ_TurbulentFragment-IMF,Hopkins_2013MNRAS_TurbulentFragment,Chabrier_2014ApJ_IMF_StarburstEnvironment,Barber_2018MNRAS_Variable_IMF1,Barber_2019MNRAS_Variable_IMF2,Barber_2019MNRAS_Variable_IMF3})

\section{Conclusions}
\label{section:6}
In this work, we study the modeling bias from constant stellar mass-to-light ratio assumption, which has been widely used in galaxy dynamics and lensing studies (e.g., \citealt{Treu_2002ApJ_ETG_MG2016+112,Koopmans_2006ApJ_SLACS_III,Sonnenfeld_2012ApJ_ETG_SalpeterIMF,Sonnenfeld_2015ApJ_SL2S_V}). We take a sample of over one hundred of simulated galaxies from the TNG100 project (\citealt{Weinberger_2017MNRAS_AGNmod,Pillepich_2018MNRAS_IllustrisTNG,Pillepich_2018MNRAS_TNG_StellarMassCotent,Nelson_2018MNRAS_TNG_GalColorBimodality,Springel_2018MNRAS_TNG_GalClustering,Naiman_2018MNRAS_TNG_ChemicalEvo,Marinacci_2018MNRAS_TNG_RadioMagnet}) and fit their total density profiles with two-component models within a Bayesian framework, following procedures widely adopted in observations (e.g., \citealt{Sonnenfeld_2015ApJ_SL2S_V,Oldham_2018MNRAS_M87_IMF,Oldham_2018MNRAS_SL,Sonnenfeld_2018MNRAS_SL_WL}). In particular, we use a generalized NFW model (\citealt{ZhaoHongsheng_1996MNRAS_gNFW,Wyithe_2001ApJ_gNFW}) to describe the dark matter halo density profile. We fit the light surface brightness profile with both a single- and double-S{\'e}rsic model (\citealt{Sersic_1963BAAA}). We then multiply the best-fit light model with a constant stellar-mass to light ratio $\Upsilon_\star$, which is then used to describe the stellar mass density model. We fit the two-component model to the total density profiles of simulated galaxies, and examined the modeling performances using the density profile data within $R_{200}$ in an all-free model (with flat prior for all model parameters, as our baseline model, see section~\ref{section:2.3.1} for details) and in a model that assumes a Gaussian prior on $\Upsilon_\star$ (centred on the true value and with a scatter of $\sim 10\%$, see section~\ref{section:2.3.2}). The other two models restrict the range to $4R_{\rm eff}$, emulating the study without constraints at large radii. In addition, some priors on the model parameters are set in models with partially constrained gNFW profiles where either the dark matter inner slope was fixed to 1 (see section~\ref{section:2.3.3}), or the halo concentration was fixed by empirical relations (see section~\ref{section:2.3.4}). 

We find that the $M_{\star}/L$ profile of simulated massive early-type sample generally declines with radius, i.e., the true stellar mass density distributions more centrally concentrated than the luminosity density distributions (see Fig.~\ref{fig:four_examples_light} and section~\ref{section:3}). In particular, we define the galaxies that profiles possess comparatively shallower logarithmic $M_\star/L$ gradient ($\alpha<0.1$, see Fig.~\ref{fig:m2l_stack_prof}) as the constant $M_\star/L$ system. However, a considerable fraction of galaxies exhibit markedly non-constant $M_{\star}/L$ profiles ($\alpha>0.1$ as the non-constant $M_\star/L$ galaxy). Almost half of our simulated massive early-type galaxies have a significant $M_\star/L$ gradient, indicating that assuming constant $M_\star/L$ is not a good approximation. The $M_\star/L$ variation of these simulated galaxies is typically around the region, where on-going star formation exists within a quenched stellar halo (see Fig.~\ref{fig:m2l_stack_prof} and Fig.~\ref{fig:four_examples_map}). This often causes $M_{\star}/L$ to decrease dramatically at the associated radii, and leads to seemingly steeper $M_\star/L$ gradient.


In previous works, the lensing and stellar kinematic data can effectively constrain the total mass distribution. But the dark matter distribution is still degenerate with baryons in the galaxy centre. We find that the assumption of a constant stellar mass-to-light ratio $M_{\star}/L$ would in general artificially break such a degeneracy and causes non-negligible model biases for those non-constant $M_\star/L$ galaxies (i.e., the model prefers under-estimating one specific component). In composite model fitting where an all-free model (baseline model) has been adopted (see section~\ref{section:4.1}), due to the similarity of the logarithmic slope between central light distribution and the total density profile, the model would artificially enhance $\Upsilon_\star$ and thus result in under-estimated $f_{\rm DM}$ and flatter $\gamma_{\rm DM}$. When we take the single-S{\'e}rsic model to describe the stellar component for those galaxies with a non-constant stellar mass-to-light ratio, the mean bias of the stellar mass-to-light ratio $\log{\Upsilon^{(\rm mod)}_{\star}/\Upsilon^{(\rm true)}_{\star}}$ is $0.19 \pm 0.07\, \rm dex$, and the mean bias of the dark matter fraction $f^{(\rm mod)}_{\rm DM}-f^{(\rm true)}_{\rm DM}$ is $-0.18 \pm 0.08$. As a comparison, for those galaxies with a constant stellar mass-to-light ratio, the mean biases as well as their errors of $\log\Upsilon_\star/\Upsilon_\odot$ and $f_{\rm DM}$ are much smaller, $-0.04 \pm 0.14$ and $0.02 \pm 0.12$, respectively. When we consider the double-S{\'e}rsic model for the stellar component, the general trend between the $M_\star/L$ and modeling biases is not significantly changed.

It is then interesting to ask what if we have some knowledge about the average stellar mass-to-light ratio $\Upsilon_\star$? To test this, we have then added a prior on $\Upsilon_\star$ centered around the truth. The model in general can better reproduce the central dark matter fraction $f_{\rm DM}$. However, as the model now cannot freely enhance $\Upsilon_\star$ to compensate for a more compact stellar mass density distribution, it is only left with the choice of systematically overestimating the dark matter inner slope $\gamma _{\rm DM}$ for those non-constant $M_\star/L$ systems (see section~\ref{section:4.2.1}). We have also partially constrained the gNFW model to describe the dark matter component only with the data from the central region. These fixed models, such as NFW (see section~\ref{section:4.2.2}) or assuming $r_{\rm DM} = 10R_{\rm eff}$ (section~\ref{section:4.2.3}) also introduce additional biases. In particular, the application of NFW model results in a systematically under-predicted dark matter component since the simulated massive galaxies always keep a $\gamma_{\rm DM}$ exceeding 1.


As can be seen, without a good description to $M_{\star}/L$, different models (on both dark matter and light) would artificially break the degeneracy in different manners, therefore resulting in different inferences. This study, investigating the consequence of the constant $M_{\star}/L$ assumption, simply builds up the first stage of a series of investigation along this direction.

\section*{Acknowledgements}

We thank Dr. Shengdong Lu for helpful and constructive discussions about this paper. We thank the anonymous referees whose comments have helped to significantly improve the quality of the paper. This work is supported by the China Manned Space Project (No. CMS-CSST-2021-A07). LY and XDD also acknowledge the General Program from National Natural Science Foundation of China (No. 12073013). We would like to thank the high-performance computing cluster at the Department of Astronomy, Tsinghua University for providing computational and data storage resources for this study.

\section*{Data Availability}

The data of simulated galaxies in this work come from the TNG100 cosmological simulation project. All of the data are public and can be obtained on the website of the TNG project (\href{https://www.tng-project.org}{https://www.tng-project.org}). We will also share the galaxy IDs and other underlying data for reasonable requests. 



\bibliographystyle{mnras}
\bibliography{example} 




\appendix

\section{The Uncertainty Estimation}
\label{section:A1}

\begin{figure}
    \centering
    \includegraphics[width=\columnwidth]{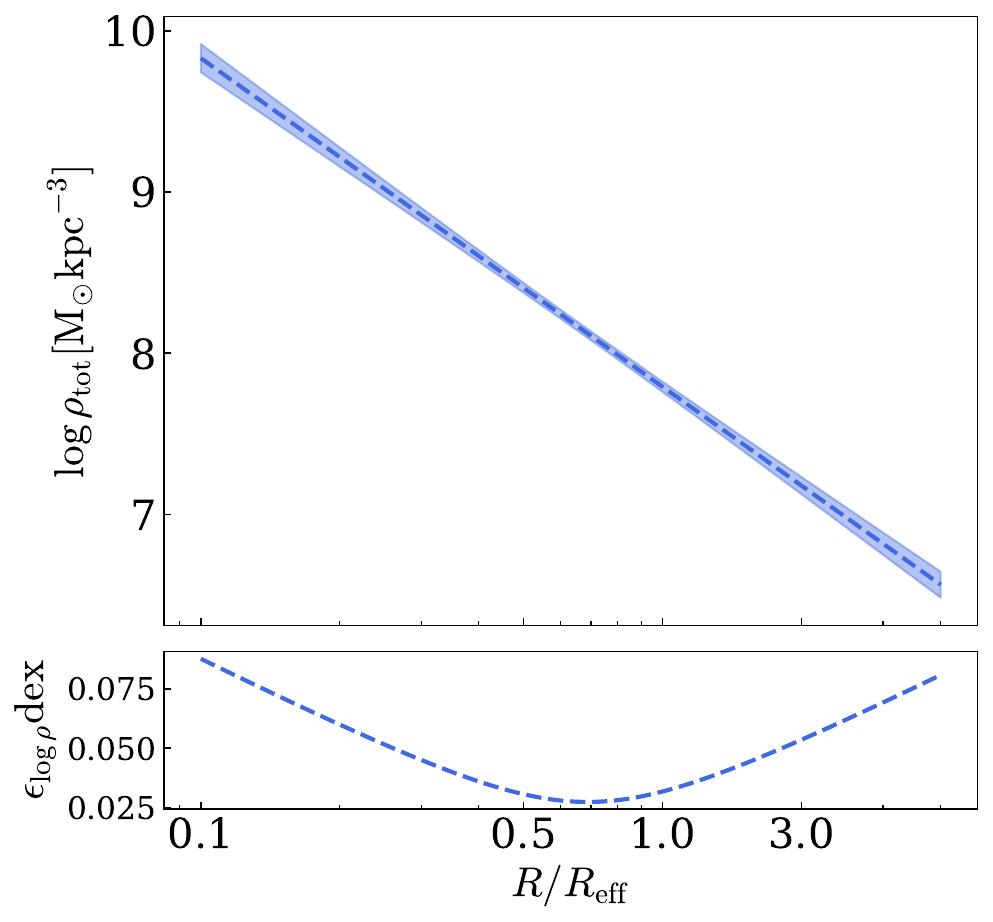}
    \caption{The total density profile and its uncertainty of our mock strong lensing galaxy.}
    \label{fig:mock_profile}
\end{figure}

In section~\ref{section:2.3}, we set the uncertainty for the logarithmic density data to be $0.1 \rm dex$. Such an uncertainty is obtained from observational perspective. Observationally, the uncertainty on Einstein mass and velocity dispersion measurement are roughly at 10 \% level (e.g. \citealt{Kochanek_1991ApJ}). The problem is that the current accuracy (typically 10\%) in the velocity dispersion measurements may not yet be sufficient to break the mass sheet degeneracy, not to mention the uncertainty due to the anisotropy of stellar orbits, which can lead to a systematic error on the slope at a level of $\sim 5\%$ (\citealt{Agnello_2013MNRAS}). To avoid this problem, we generate a mock lensing galaxy. We use the average Einstein radius and average velocity dispersion of the lensing galaxy population in SL2S project (\citealt{Sonnenfeld_2013ApJ_SL2S_III, Sonnenfeld_2013ApJ_SL2S_IV}) as the observables of our mock lensing galaxy. And the uncertainties of this mock lensing's observable are also the average uncertainties of Einstein radii and velocity dispersions of the lensing galaxies in SL2S project. We assume a spherical power law density distribution for this mock lensing system and a de Vaucouleurs model for the surface brightness profile. After that, we combine the strong lensing and the spherical Jeans modeling to infer the total density of this mock galaxy with the Bayesian framework. Finally, we resample 1000 total density model parameter pairs (the total mass within Einstein radius and total density power law slope) from their posterior distributions and get 1000 possible total density profiles of our mock galaxy. The result is shown in Fig.~\ref{fig:mock_profile} in which the upper panel is the total density profile while the bottom panel is the standard deviation at different radii. We find that $\epsilon_{\log{\rho}}$ is not larger than $0.08 \rm dex$ from $0.1 R_{\rm eff}$ to $4R_{\rm eff}$. Since we only consider the strong lensing and stellar kinematic constraint in this mock system, we still need to estimate the observational uncertainty total density approximately up to virial radius where the weak lensing is widely used to measure the mass of galaxies or galaxy clusters. According to previous weak lensing studies (\citealt{Hoekstra_2004ApJ_DMhalo_WL,Hoekstra_2005ApJ_WL_Lensing}), the observational uncertainty on the virial mass is roughly $0.1 \rm dex$. Since the slope of the galaxy total density at large radii is dominated by the dark matter halo and fixed in our dark matter model, the uncertainty of the virial mass approximates to the uncertainty of the total density at large radii. Without loss of generality, we set the uncertainty of total density profile data in this work to $0.1 \rm dex$        

\section{The Results of the Entire Early-Type Sample}
\label{section:B1}
\begin{figure*} 
    \centering
    \includegraphics[width=\textwidth]{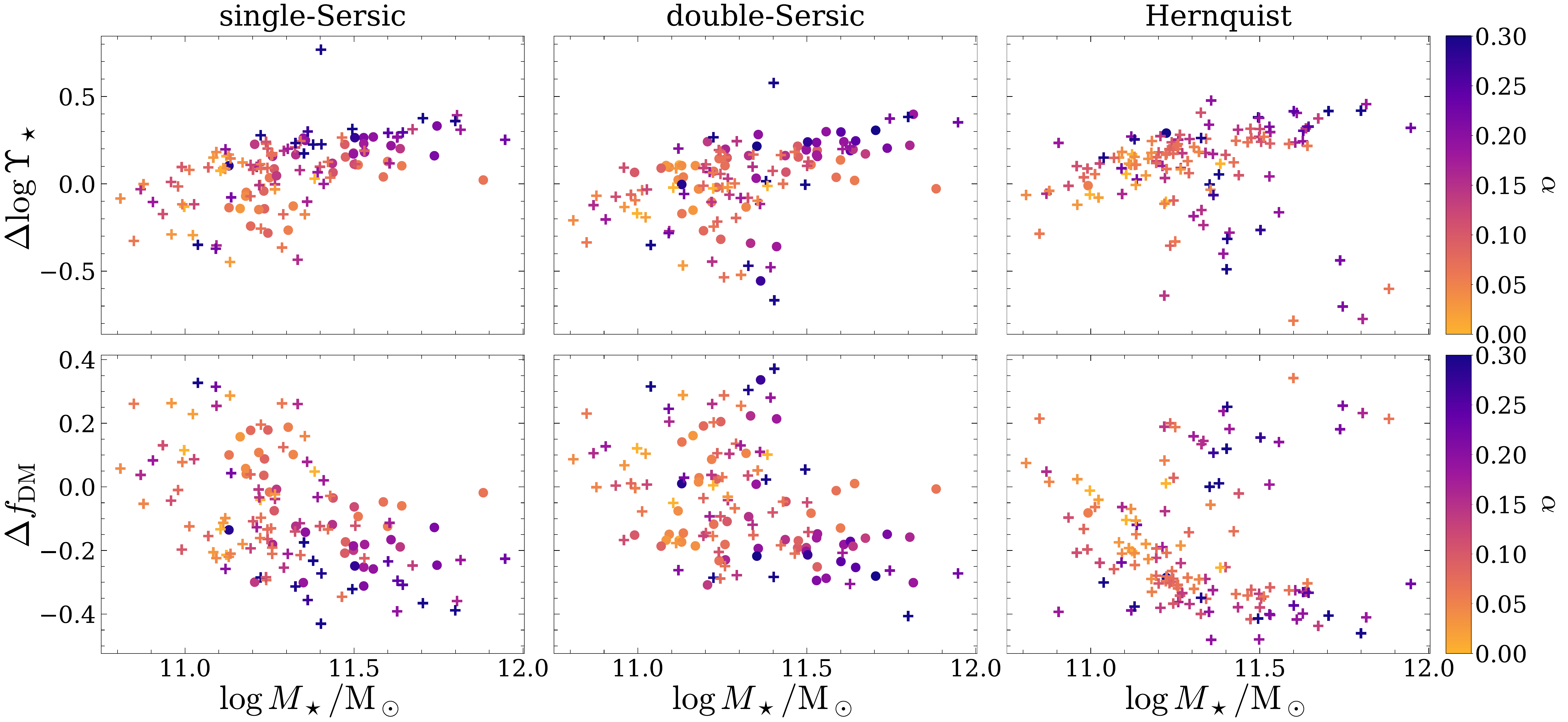}
    \caption{The modeling biases as the function of total stellar mass of complete massive early-type sample for all-free scenario. Each column represents one individual light model. As we have shown in Fig.~\ref{fig:bias_alpha} and Fig.~\ref{fig:results_alter}, $\Delta\log\Upsilon_\star/\Upsilon_\odot$ and $\Delta f_{\rm DM}$ are the difference between the model predicted value and the ground truth. All data points are color-coded by the $M_\star/L$ gradient. The circular points belong to the sub-sample whose light distribution can be well described by our models. The crosses are the remaining massive early-type galaxies, which cannot satisfy our further criteria about light model.}
    \label{fig:results_all}
\end{figure*}

In section~\ref{section:4}, we have presented and discussed the modeling biases under the constant $M_\star/L$. Our analysis is confined to galaxies whose light profiles are well captured by our light models, specifically the single- and double-S{\'e}rsic profiles. Moreover, we test the possibility of utilizing the Hernquist model for the stellar component, yet the number of systems fulfilling our further criteria for the Hernquist light model are insufficient to support a robust statistical analysis. Consequently, the results based on adopting Hernquist profiles are excluded from section ~\ref{section:4}. In this appendix, we present the fitting results of the entire massive early-type galaxy sample, applying all three light models. Fig.~\ref{fig:results_all} shows the inference biases as a function of total stellar mass for all 129 massive early-type galaxies. We find that the general trend of the modeling bias for single- or double-S{\'e}rsic model is not changed, the galaxy with a comparatively steep $M_\star/L$ gradient has a more biased inference (enhanced stellar component). But a considerable fraction of galaxies without good description from light model exhibit the opposite prediction, characterized by an overestimation of the dark matter component despite having $\alpha>0.1$ (especially for the Hernquist light model). We emphasis that such a contrasting inference is attributed to both $M_\star/L$ and model incompatibility.


\bsp	
\label{lastpage}
\end{document}